\documentclass{sig-alternate-05-2015}

\usepackage{graphicx}
\usepackage{color}
\usepackage{algorithm}
\usepackage[noend]{algorithmic}
\usepackage{varwidth}
\usepackage{capt-of}
\usepackage{balance}  
\usepackage{times,amsmath,epsfig}
\usepackage{enumitem}

\newcommand{\eat}[1]{}
\newcommand{\A}[1]{#1} 

\newcommand{\E}[1]{\emph{#1}}
\newcommand{\R}[1]{\textbf{#1}}

\newcommand{\emptySet}{\brac{}}
\newcommand{\brac}[1]{\{#1\}}

\newcommand{\eatTR}[1]{#1} 

\newcommand{\eatNTR}[1]{} 

\newcommand{\set}[1]{#1}

\newtheorem{example}{Example}
\newtheorem{definition}{Definition}
\newtheorem{theorem}{Theorem}
\newtheorem{lemma}[theorem]{Lemma}

\newfont{\mycrnotice}{ptmr8t at 7pt}
\newfont{\myconfname}{ptmri8t at 7pt}
\let\crnotice\mycrnotice%
\let\confname\myconfname%

\begin{document}


\pdfoutput=1

\title{Efficient Discovery of Ontology Functional Dependencies}

\author{
\alignauthor
Sridevi Baskaran$\small^{^{*}}$, Alexander Keller$\small^{^{\#}}$, Fei Chiang$\small^{^{*}}$, Lukasz Golab$\small^{^{\$}}$, 
    Jaroslaw Szlichta$\small^{^{\#}}$\\
        \affaddr{$^{*}$\normalsize McMaster University, Canada}\\
        \affaddr{$^{\#}$\normalsize University of Ontario Institute of Technology, Canada}\\
        \affaddr{$^{\$}$\normalsize University of Waterloo, Canada}\\
        \affaddr{\normalsize baskas@mcmaster.ca, alexander.keller@uoit.net, fchiang@mcmaster.ca, lgolab@uwaterloo.ca, jaroslaw.szlichta@uoit.ca}
}

\maketitle
\begin{abstract}
Functional Dependencies (FDs) define attribute relationships based on syntactic equality, and, when used in data cleaning, they erroneously label syntactically different but semantically equivalent values as data errors. In this paper, we enhance dependency-based data cleaning with \emph{Ontology FDs} (OFDs), which express semantic attribute relationships such as synonyms and is-a hierarchies defined by an ontology.  Our technical contributions are twofold: 1) theoretical foundations for OFDs, including a set of sound and complete axioms and a linear-time inference procedure, and 2) an algorithm for discovering OFDs (exact ones and ones that hold with some exceptions) from data that uses the axioms to prune the exponential search space.  We demonstrate the efficiency of our techniques on real datasets, and we show that OFDs can significantly reduce the number of false positive errors in data cleaning techniques that rely on traditional FDs.
\end{abstract}

\maketitle
\section{Introduction}
\label{sec:intro}
Organizations are finding it increasingly difficult to reap value from their data due to poor data quality.  The increasing size and complexity of modern datasets exacerbate the fact that real data is dirty, containing inconsistent, duplicated, and missing values.  A Gartner Research study reports that by 2017, 33\% of the largest global companies will experience a data quality crisis due to their inability to trust and govern their enterprise information \eatTR{\cite{JF14}}\eatNTR{\cite{JF14}}. 

In constraint-based data cleaning, dependencies are used to specify data quality requirements.  Data that are inconsistent with respect to the dependencies are identified as erroneous, and updates to the data are generated to re-align the data with the dependencies.  Deciding which updates to apply often goes beyond simply identifying similar or equivalent string values, as user input is necessary to gather information such as terminologies, concepts, and relationships that are relevant for a given application domain.

Existing data cleaning approaches have focused on using functional dependencies (FDs) to define the attribute relationships that the data must satisfy \cite{BFFR05,PSC15}.  Extensions include the use of inclusion dependencies \cite{BFFR05}, conditional functional dependencies \cite{CFGJM07}, and denial constraints \cite{CIP13}.  However, these approaches
are limited to identifying attribute relationships based on syntactic equivalence (or syntactic similarity in case of Metric FDs \cite{KSS09,PSC15}), as shown in the following example.

\begin{table*}[ht]
\begin{small}

 \begin{varwidth}[b]{0.45\linewidth}
\centering
\begin{tabular}{ | l | l | l | l | l | l |}
\hline
\hspace{-2 mm} \textbf{id} \hspace{-2 mm} & \textbf{CC}    & \textbf{CTRY} & \hspace{-2 mm} \textbf{SYMP} & \hspace{-2 mm} \textbf{DIAG}  & \hspace{-2 mm} \textbf{MED}  \\
\hline \hline
\hspace{-2 mm} $t_1$ \hspace{-2 mm} &      US & United States &  joint pain & osteoarthritis & ibuprofen \\
\hspace{-2 mm} $t_2$ \hspace{-2 mm} &      IN & India & joint pain & osteoarthritis & NSAID  \\
\hspace{-2 mm} $t_3$ \hspace{-2 mm} &     CA & Canada & joint pain & osteoarthritis & naproxen \\
\hspace{-2 mm} $t_4$ \hspace{-2 mm} &      IN & Bharat & nausea & migrane &  analgesic  \\
\hspace{-2 mm} $t_5$ \hspace{-2 mm} &       US & America & nausea \ & migrane &  tylenol \\
\hspace{-2 mm} $t_6$ \hspace{-2 mm} &      US & USA &  nausea & migrane &  acetaminophen \\
\hspace{-2 mm} $t_7$ \hspace{-2 mm} &       IN & India & chest pain & hypertension & morphine   \\
\hline
\end{tabular}
\caption{Sample clinical trials data}
\vskip -0.5cm
\label{tab:example}
\end{varwidth}%
\hfill
\begin{minipage}[b]{0.19\linewidth}
  \centering
\includegraphics[width=1.8in]{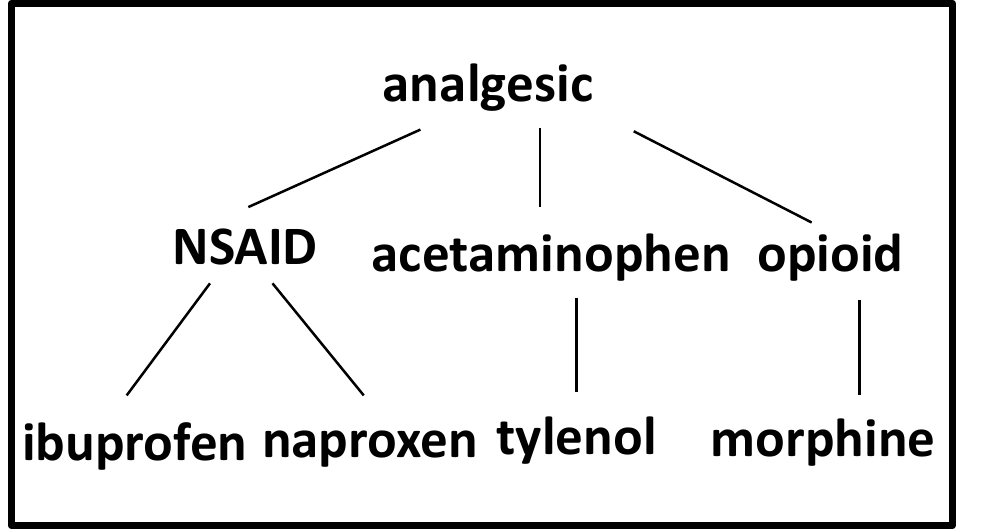}\\
\captionof{figure}{An ontology}
\vskip -0.05cm
\label{fig:ontology}
\end{minipage}
\hfill
\begin{varwidth}[b]{0.2\linewidth}
\begin{small}
\centering
   \begin{tabular}{|c||c|c||c|}
       \hline
        id & X & Y & Classes for Y\\
       \hline \hline
       $t_{1}$ &  \emph{u} &  \emph{v} &  \{\emph{C,D}\}\\
       \hline
      $t_{2}$  &  \emph{u} & \emph{w} &   \{\emph{D,F}\}\\
       \hline
      $t_{3}$  &  \emph{u} &  \emph{z} &  \{\emph{C,F,G}\}\\
       \hline
   \end{tabular}
    \vskip 1.2cm
    \caption{Defining OFDs}
    \vspace{-0.5cm}
   \label{tab:mVio}
\end{small}
\end{varwidth}

\end{small}
\end{table*}

\begin{example}
\label{ex:intro}
Table \ref{tab:example} shows a sample of clinical trial records containing patient country codes (CC), country (CTRY), symptoms (SYMP),  diagnosis (DIAG), and the prescribed medication (MED).   Consider two FDs: $F_{1}$:  [CC]  $\rightarrow$  [CTRY] and $F_{2}$: [SYMP, DIAG] $\rightarrow$ [MED].   The tuples ($t_{1}, t_{5}$, $t_{6}$) do not satisfy $F_{1}$ as `United States', `America', and `USA' are \emph{not syntactically (string) equivalent} (the same is true for ($t_{2}, t_{4}$, $t_{7}$)).   However, `United States' is synonymous with `America' and `USA', and ($t_{1}, t_{5}$, $t_{6}$) all refer to the same country.  Similarly, `Bharat' in $t_{4}$ is synonymous with `India' as it is the country's original Sanskrit name.   For $F_{2}$, ($t_{1}, t_{2}, t_{3}$) and ($t_{4}, t_{5}, t_{6}$) do not satisfy the dependency as the consequent values all refer to different medications.  However, with domain knowledge from a medical ontology (Figure \ref{fig:ontology}), we see that the values participate in an inheritance relationship.  Both `ibuprofen' and `naproxen' are non-steroidal anti-inflammatory drugs (NSAID), and `tylenol' is an `acetaminophen' drug, which in turn is an `analgesic'.
\end{example}

The above example demonstrates that real data contain domain-specific relationships that go beyond syntactic equivalence or similarity.  It also highlights two common relationships that occur between two values $u$ and $v$: (1) $u$ and $v$ are \emph{synonyms}; and (2) $u$ is-a $v$ denoting \emph{inheritance}.  These relationships are often defined within domain specific ontologies that can be leveraged during the data cleaning process to identify and enforce domain specific data quality rules.  Unfortunately, traditional FDs and their extensions are unable to capture these relationships, and existing data cleaning approaches flag tuples containing synonymous and inheritance values as errors.  This leads to an increased number of "errors", and a larger search space of data repairs to consider.

To address these shortcomings, we study a novel class of dependencies called \emph{Ontology Functional Dependencies} (OFDs) that capture synonyms and is-a relationships defined in an ontology.  As we will show, what makes OFDs non-trivial and interesting is the notion of \emph{senses}, which determine how a dependency should be interpreted for a given ontology; e.g., 'jaguar' can be interpreted as an \emph{animal} or as a \emph{vehicle}.  Notably, interpretations have not been considered in prior work on database or ontology quality (see Section~\ref{sec:rw} for a detailed discussion of prior work).  Furthermore, to make OFDs useful in practice, where data semantics are often poorly documented and change over time, we propose an algorithm to discover OFDs from data.  
We make the following contributions.

\begin{enumerate}[nolistsep,leftmargin=*]
\item We define OFDs based on the synonym and inheritance relationships.  In contrast to existing work, OFDs include  attribute relationships that go beyond equality, and consider the notion of \emph{senses} that provide the interpretations under which the dependencies are evaluated.  

\item We introduce a \emph{sound and complete} set of axioms for OFDs.  While the inference complexity of other FD extensions is co-NP complete, we show that the inference problem for OFDs remains linear.

\item
We propose an algorithm to discover a \emph{complete} and \emph{minimal} set of OFDs from data in order to alleviate the burden of specifying them manually for data cleaning.  We show that 
OFDs can be discovered efficiently by traversing a \emph{set-containment lattice} with \emph{exponential} worst-case complexity in the number of attributes, the same as for traditional FDs~\cite{HKPT98}, and polynomial complexity in the number of tuples.  
%
We develop optimizations based on our axiomatization which prune the search space without affecting correctness.
%
%
\item We introduce \emph{approximate} OFDs and show they are a useful data quality rule to capture domain relationships, and can significantly reduce the number of false positives in data cleaning techniques that rely on traditional FDs.
\item We evaluate the performance and effectiveness of our techniques using two real datasets containing up to 1M records.  
\end{enumerate}

We present preliminaries and definitions in Section~\ref{sec:prelim}.  
Section \ref{sec:foundations} presents our axiomatization and inference procedure for OFDs, and Section \ref{sec:fastofd} describes our OFD discovery algorithm.
We discuss experimental results in Section \ref{sec:eval}, related work in Section \ref{sec:rw}, and directions for future work in Section \ref{sec:conclusion}.

\section{Preliminaries and Definitions}
\label{sec:prelim}

A functional dependency (FD) $F$ over a relation $R$ is represented as $X \rightarrow A$, where $X$ is a set of attributes and $A$ is a single attribute in $R$.  An instance $I$ of $R$ satisfies $F$ if for every pair of tuples $t_1, t_2 \in I$, if $t_1$[$X$] = $t_2$[$X$], then $t_1$[$A$] = $t_2$[$A$].  A partition of $X$, $\Pi_{X}$, is the set of equivalence classes containing tuples with equal values in $X$.  Let $x_i$ be an equivalence class with a representative $i$ that is equal to the smallest tuple id in the class, and $|x_i|$ be the size of the equivalence class (in some definitions, we drop the subscript and refer to an equivalence class simply as $x$).  For example, in Table \ref{tab:example}, $\Pi_{CC}$ = \{\{$t_1,t_5,t_6$\}\{$t_2,t_4,t_7$\}\{$t_3$\}\}.  

An ontology $S$ is an explicit specification of a domain that includes concepts, entities, properties, and relationships among them.  These constructs are often defined and applicable only under a given interpretation, called a \emph{sense}.   The meaning of these constructs for a given $S$ can be modeled according to different senses leading to different ontological interpretations.  As mentioned previously, the value `\texttt{jaguar}' can be interpreted under two senses: (1) as an animal, and (2) as a vehicle.  As an animal, `\texttt{jaguar}' is synonymous with `\texttt{panthera onca}', but not with the value `\texttt{jaguar land rover}' which is an automotive company.  

We define classes $E$ for the interpretations or senses defined in $S$.
Let \emph{synonyms}$(E)$ be the set of all synonyms for a given class $E$. For instance, $synonyms(E1)$ = \{\emph{'car', 'auto', 'vehicle'}\}, $synonyms(E2)$ = \{\emph{'jaguar', 'jaguar land rover'}\} and $synonyms(E3)$ = \{\emph{'jaguar', 'panthera onca'}\}. Let \emph{names}$(C)$ be the set of all classes, i.e., interpretations or senses, for a given value $C$. For example, \emph{names}$(jaguar)$ = \{$E2$, $E3$\} as jaguar can be an animal or a vehicle. Let $descendants(E)$ be a set of all string representations for the class $E$ or any of its descendants, i.e., $descendants(E)$ = \{$s$ $|$ $s$ $\in$ \emph{synonyms}$(E)$  or $s$ $\in$ \emph{synonyms}$(E_{i})$, where $E_{i}$ \emph{is-a} $E_{i-1}$, ..., $E_{1}$ \emph{is-a} $E_{0}$\}. For instance, $descendants(E3)$ = \{\emph{'jaguar', 'panthera onca', 'feline', 'mammal', 'animal'}\}.

We define OFDs with respect to a given ontology $S$.  We consider two relationships, synonyms and inheritance, leading to synonym OFDs (Definition~\ref{def:synonym_ofd}) and inheritance OFDs (Definition~\ref{def:inheritance_ofd}). 

\begin{definition} \label{def:synonym_ofd}
A relation instance $I$ satisfies a \emph{synonym OFD} $X \rightarrow_{syn} A$, if for each equivalence class $x \in \Pi_{X}(I)$, there exists an interpretation under which all the $A$-values of tuples in $x$ are synonyms.  That is, $X \rightarrow_{syn} A$ holds if for each equivalence class $x$, $$\bigcap_{names(a), a \in \{t[A] | t \in x \}} \neq \emptyset.$$
\end{definition}

\begin{example}\label{example:syn}
Consider the OFD [CC]  $\rightarrow_{syn}$  [CTRY]  from Table \ref{tab:example}.  We have 
$\Pi_{CC}$ = \{\{$t_1,t_5,t_6$\}\{$t_2,t_4,t_7$\}\{$t_3$\}\}.  The first equivalence class, \{$t_1,t_5,t_6$\}, representing the value `US', corresponds to three distinct values of CTRY.  According to a geographical ontology, names(`United States') $\cap$ names(`America') $\cap$ names(`USA') = `United States of America'.  
Similarly, the second class \{$t_2,t_4,t_7$\} gives names(`India') $\cap$ names(`Bharat') = `India'.  The last equivalence class \{$t_3$\} contains a single tuple, so there is no conflict.  Since all references to CTRY in each equivalence class resolve to some common interpretation,  
the OFD holds over Table \ref{tab:example}.
\end{example}

By definition, synonym OFDs subsume traditional FDs, where all values are assumed to have a single literal interpretation (for all classes $E$, $|$\emph{synonyms}$(E)|$ = 1). 

\begin{definition}  \label{def:inheritance_ofd}
Let $\theta$ be a threshold representing the allowed path length between two nodes in $S$.  A relation instance $I$ satisfies an \emph{inheritance OFD} $X \rightarrow_{\theta} Y$ if for each equivalence class $x \in \Pi_{X}(I)$, the $A$-values of all tuples in $x$ are descendants of a Least Common Ancestor (LCA) which is within a distance of $\theta$ in $S$.  That is, $X \rightarrow_{\theta} A$ holds if for each equivalence class $x$, $$\bigcap_{descendants(names(a)), a \in \{t[A] | t \in x \}} \neq \emptyset$$ and the resulting LCA is within a distance of $\theta$ in $S$ to each $a$.
\end{definition}



\begin{example}\label{example:isa}
Consider the OFD [SYMP, DIAG] $\rightarrow_{\theta}$ [MED] from Table \ref{tab:example} and the ontology in Figure \ref{fig:ontology}.  For the first equivalence class, $\{t_1,t_2,t_3\}$, the LCA is `NSAID' which is within a distance of one to each MED value in this class.  For the second equivalence class, $\{t_4,t_5,t_6\}$, the LCA is `analgesic', which is within a distance of two to each MED value in this class.  The third equivalence class consists of a single tuple ($t_7$) so there is no conflict.  Thus, [SYMP, DIAG] $\rightarrow_{2}$ [MED] holds.
\end{example}

The threshold $\theta$ ensures that values deemed semantically equivalent are sufficiently similar; clearly, if we search high enough in $S$, we can always find a common ancestor. In practice, users should set $\theta$ based on the granularity of the associated ontology.

By definition, inheritance OFDs subsume synonym OFDs since we can recover synonym OFDs by setting $\theta = 0$. For example, the inheritance OFD [SYMP, DIAG] $\rightarrow_{0}$ [MED] allows synonyms such as  'ibuprofen' and 'advil' to appear in tuples having the same SYMP and DIAG values; however, different medications under the same LCA such as 'ibuprofen' and 'naproxen' are not allowed. 

The notion of senses makes OFDs non-trivial and interesting.  For each equivalence class, there must exist a common interpretation of the $Y$-values. Checking pairs of tuples, as in traditional FDs (and Metric FDs \cite{KSS09,PSC15}, which assert that two tuples whose left-hand side attribute values are equal must have \emph{syntactically} similar right-hand side attribute values according to some distance metric), is not sufficient, as illustrated below.

\begin{example}
Consider Table \ref{tab:mVio}, where the synonym OFD $X \rightarrow_{syn} Y$ does not hold (for each $Y$ value, we list its possible interpretations in the last column). 
Although all pairs of $t[Y]$ values share a common class (i.e., \{\emph{v,w}\}: \E{D}, \{\emph{v,z}\}: \E{C}, \{\emph{w,z}\}: \E{F}), the intersection of the classes is empty. 
\end{example}


Note that OFDs \emph{cannot} be reduced to traditional FDs or Metric FDs.  Since values may have multiple senses (e.g., jaguar the animal and jaguar the car), it is not possible to create a normalized relation instance by replacing each value with a unique canonical name.  Furthermore, ontological similarity is not a metric since it does not satisfy the identity of indiscernibles (e.g., for synonyms).

\textbf{Problem Statement:} Given a relational instance $I$ and a threshold $\theta$, we want to discover a complete, minimal (non-redundant) set of synonym and inheritance OFDs.  $X \rightarrow A$ is \emph{trivial} if $A \in X$, and $X \rightarrow A$ is minimal if there is no $Z \rightarrow A$ that holds such that $Z \subset X$. 

\section{Foundations and Optimizations}
\label{sec:foundations}

In this section, we provide a formal framework for OFDs.  We give a sound and complete axiomatization for OFDs which reveals how OFDs behave; notably, not all axioms that hold for traditional FDs carry over.  We then use the axioms to design pruning rules that will be used by our OFD discovery algorithm.
Finally, we provide a linear time inference procedure that ensures a set of OFDs remains minimal.   
\eatNTR{Due to space constraints, we defer all proofs to the accompanying technical report~\cite{BKCS16}.}

\subsection{Axiomatization for OFDs}


We start with the \emph{closure} of a set of attributes $\set{X}$  over a set of OFDs $\set{M}$, which will allow us to determine whether additional OFDs follow from $\set{M}$ by axioms. We use the notation $\set{M} \vdash$ to state that $\set{X}$ $\rightarrow$ $\set{Y}$ is provable with axioms from $\set{M}$. 

\begin{definition} $(closure)$ \label{def:closure}
The closure of $\set{X}$, denoted as $\set{X}^{+}$, with respect to the set of OFDs $M$ is defined as $\set{X}^{+}$ $=$ $\{\A{A}$ $|$ $\set{M}$ $\vdash$ $\set{X}$ $\rightarrow$ $\A{A} \}$.
\end{definition}

\eatTR{\begin{lemma}\label{lemma:closure}
$\set{M}$ $\vdash$ $\set{X}$ $\rightarrow$ $\set{Y}$ iff $\set{Y}$ $\subseteq$ $\set{X}^{+}$.
\end{lemma}}

\eatTR{\begin{proof}
Let $\set{Y}$ $=$ $\{\A{A}_{1}$, $...$, $\A{A}_{n}\}$. Assume $\set{Y}$ $\subseteq$ $\set{X}^{+}$. By definition of $\set{X}^{+}$, $\set{X}$ $\rightarrow$ $\set{A}_{i}$, for all $i$ $\in$ $\{1,..., n\}$. Therefore, by Union inference rule, $\set{X}$ $\rightarrow$ $\set{Y}$ follows. The other direction, suppose $\set{X}$ $\rightarrow$ $\set{Y}$ follows from the axioms. For each $i$ $\in$ $\{1,..., n\}$, $\set{X}$ $\rightarrow$ $\A{A}_{i}$ follows by the Decomposition axiom. Therefore, $\set{Y}$
$\subseteq$ $\set{X}^{+}$.
\end{proof}}

Theorem~\ref{theorem:complete} presents a sound a complete set of axioms (inference rules) for OFDs. The Identity axiom generates \emph{trivial} dependencies, which are always true.

\begin{theorem}\label{theorem:complete}
These axioms are sound and complete for OFDs.
    \begin{enumerate}[nolistsep]
        \item Identity: $\forall X \subseteq R$, $X$ $\rightarrow$ $X$
        \item Decomposition: If $X$ $\rightarrow$ $Y$ and $Z$ $\subseteq$ $Y$
            then $X$ $\rightarrow$ $Z$
        \item Composition: If $X$ $\rightarrow$ $Y$ and $Z$ $\rightarrow$
        $W$ then $XZ$ $\rightarrow$ $YW$
    \end{enumerate}
\end{theorem}

\eatTR{\begin{proof}
First we prove that the axioms are sound. That is, if $\set{\set{M}}$ $\vdash$ $\set{X}$ $\rightarrow$ $\set{Y}$, then $\set{M}$ $\models$ $\set{X}$ $\rightarrow$ $\set{Y}$. The Identity axiom is clearly sound. We cannot have a relation with tuples that agree on $\set{X}$ yet are not in synonym or inheritance relationship, respectively. To prove Decomposition, suppose we have a relation that satisfies $\set{X}$ $\rightarrow$ $\set{Y}$ and $\set{Z} \subseteq \set{Y}$. Therefore, for all tuples that agree on $\set{X}$, they are in synonym or inheritance relationship on all attributes in $\set{Y}$ and hence, also on $\set{Z}$. Therefore, $\set{X}$ $\rightarrow$ $\set{Z}$. The soundness of Composition is an extension of the argument given previously.

Below we present the completeness proof, that is, if $\set{M}$ $\models$ $\set{X}$ $\rightarrow$ $\set{Y}$, then $\set{M}$ $\vdash$ $\set{X}$ $\rightarrow$ $\set{Y}$). Without a loss of generality, we consider a table $I$ with three tuples shown in Table \ref{table:tablet}. We divide the attributes of a relation $I$ into three subsets: $\set{X}$, the set consisting of attributes in the closure $\set{X}^{+}$ minus attributes in $\set{X}$ and all remaining attributes. Assume that the values $v$, $v'$ and $v''$ are not equal ($v$ $\not =$ $v'$, $v$ $\not =$ $v''$ and $v'$ $\not =$ $v''$), however, they are in synonym or inheritance relationship, respectively. Also, $v$, $w$ and $u$ are not in synonym nor inheritance relationship, and hence, they are also not equal.

\begin{table}[h]
\centering
\begin{tabular}{|c|c|c|}

    \hline
    \multicolumn{2}{|c|}{$\set{X}^{+}$} & \multicolumn{1}{|c|}{}\\
    \hline
    $\set{X}$ & $\set{X}^{+} \setminus \set{X}$ & Other attributes\\
    \hline
    $v...v$ & $v...v$ & $v...v$\\
    \hline
    $v...v$ & $v'...v'$ & $w...w$\\
    \hline
    $v...v$ & $v''...v''$ & $u...u$\\
    \hline
\end{tabular}
\caption{Table template for OFDs.}\label{table:tablet}
\vskip -0.3cm
\end{table}

We first show that all dependencies in the set of OFDs $M$ are satisfied by a table $I$ ($I$ $\models$ $F$). Since OFD axioms are sound, OFDs inferred from $\set{M}$ are true. Assume $\set{V}$ $\rightarrow$ $Z$ is in $\set{M}$, however, it is not satisfied by a relation $I$. Therefore,  $\set{V}$ $\subseteq$ $\set{X}$ because otherwise tuples of $I$ disagree on some attribute of $\set{V}$ since $v$, $v'$ and $v''$  as well as $v, w, u$ are not equal, and consequently an OFD $\set{V}$ $\rightarrow$ $\set{Z}$ would not be violated. Moreover, $\set{Z}$ cannot be a subset of $\set{X}^{r}$ ($\set{Z}$ $\not \subseteq$ $\set{X}^{+}$), or else $\set{V}$ $\rightarrow$ $\set{Z}$, would be satisfied by a table $I$. Let $\A{A}$ be an attribute of $\set{Z}$ not in $\set{X}^{+}$. Since, $\set{V}$ $\subseteq$ $\set{X}$, $\set{X}$ $\rightarrow$ $\set{V}$ by Reflexivity. Also a dependency $\set{V}$ $\rightarrow$ $\set{Z}$ is in $\set{M}$, hence, by Decomposition, $\set{V}$ $\rightarrow$ $\A{A}$. By Composition $\set{XV}$
$\rightarrow$ $\set{V}\A{A}$ can be inferred, therefore, $\set{X}$ $\rightarrow$ $\set{V}\A{A}$ as $\set{V}$ $\subseteq$ $\set{X}$. However, then Decomposition rule tells us that $\set{X}$ $\rightarrow$ $\A{A}$,
which would mean by the definition of the closure that $\A{A}$ is in $\set{X}^{+}$, which we assumed not to be the case. Contradiction. An OFD $\set{V}$ $\rightarrow$ $\set{Z}$ which is in $\set{M}$ is satisfied by $I$.

Our remaining proof obligation is to show that any OFD not inferable from a set of OFDs $\set{M}$ with OFD axioms ($\set{M}$ $\not \vdash$ $\set{X}$ $\rightarrow$ $\set{Y}$) is not true ($\set{M}$ $\not \models$ $\set{X}$ $\rightarrow$ $\set{Y}$). Suppose it is satisfied
($\set{M}$ $\models$ $\set{X}$ $\rightarrow$ $\set{Y}$). By Reflexivity $\set{X}$ $\rightarrow$ $\set{X}$, therefore, by Lemma \ref{lemma:closure} $\set{X}$ $\subseteq$ $X^{+}$. Since $\set{X}$ $\subseteq$ $\set{X}^{+}$ it follows by the construction of table $I$ that $\set{Y}$ $\subseteq$
$\set{X}^{+}$. Otherwise, tuples of table $I$ agree on $\set{X}$ but are not in synonym nor inheritance relationship, respectively, on some attribute $\A{A}$ from $\set{Y}$. Then, from Lemma \ref{lemma:closure} it can be inferred that $\set{X}$ $\rightarrow$ $\set{Y}$. Contradiction. Thus, whenever $\set{X}$ $\rightarrow$ $\set{Y}$ does not follow from $\set{M}$ by OFDs
axioms, $\set{M}$ does not logically imply $\set{X}$ $\rightarrow$ $\set{Y}$. That is the axiom system is complete over OFDs, and ends the proof of Theorem \ref{theorem:complete}.
\end{proof}}

A sound and complete axiomatization for traditional FDs consists of Transitivity (if $X$ $\rightarrow$ $Y$ and $Y$ $\rightarrow$ $Z$ then $X$ $\rightarrow$ $Z$), Reflexivity (if $Y \subseteq X$ then $X$ $\rightarrow$ $Y$) and Composition.  However, Transitivity does not hold for OFDs, as shown below.

\begin{example}\label{example:trans}
Consider the relation with three tuples in Table~\ref{table:transitivity}.  The synonym OFD $\A{CTRY}$ $\rightarrow_{syn}$ $\A{CC}$ holds since ``CAD'' and ``CA'' are synonyms. In addition, $\A{CC}$ $\rightarrow_{syn}$ $\A{SYMP}$ holds as ``CAD'' and ``CA'' are not equal, and ``fever" and ``pyrexia" are synonyms.  
However, the transitive synonym OFD: $\A{CTRY}$ $\rightarrow_{syn}$ $\A{SYMP}$ does not hold as ``congestion'' is not a synonym to ``fever'' nor ``pyrexia''.
\end{example}

\begin{table}[t]
\begin{center}
			\begin{tabular}{|c|c|c|c|}
				\hline
				\textbf{Patient ID} & \textbf{CTRY} & \textbf{CC} & \textbf{SYMP} \\
				\hline
				10 & Canada & CAD & fever \\
				11 & Canada & CA & congestion \\
				12 & Canada & CAD & pyrexia \\
				\hline
			\end{tabular}
		\end{center}
        
\caption{An example showing that OFDs are not transitive.}\label{table:transitivity}
\vskip -0.7cm
\end{table}


\eatTR{
\begin{lemma}$(Reflexivity)$\label{lemma:reflexivity}
If $\set{Y}$ $\subseteq$ $\set{X}$, then $\set{X}$ $\rightarrow$ $\set{Y}$.
\end{lemma}}

\eatTR{\begin{proof}
$\set{X}$ $\rightarrow$ $\set{X}$ holds by Identity axiom. Therefore, it can be inferred by the Decomposition inference rule that $\set{X}$ $\rightarrow$ $\set{Y}$ holds.
\end{proof}}

\eatTR{Union inference rule shows what can be inferred from two or more dependencies which have the same sets on the left side.}

\eatTR{\begin{lemma}$(Union)$ \label{lemma:union}
If $\set{X}$ $\rightarrow$ $\set{Y}$ and $\set{X}$ $\rightarrow$ $\set{Z}$, then $\set{X}$ $\rightarrow$ $\set{YZ}$.
\end{lemma}}

\eatTR{\begin{proof}
We are given $\set{X}$ $\rightarrow$ $\set{Y}$ and $\set{X}$ $\rightarrow$ $\set{Z}$. Hence, the Composition axiom can be used to infer $\set{X}$ $\rightarrow$ $\set{YZ}$.
\end{proof}}

\subsection{Optimizations}

As we will show in Section~\ref{sec:fastofd}, the search space of potential OFDs is exponential in the number of attributes, as with traditional FDs.  To improve the efficiency of OFD discovery, we now show how to prune redundant and non-minimal OFDs using our axiomatization.
In Section \ref{sec:opt}, we show that these optimizations significantly improve the runtime of our OFD discovery algorithm.

\begin{lemma} (Optimization 1) \\
If $A$ $\in$ $X$, then $X \rightarrow A$.
\vspace{-0.2cm}
\end{lemma}
\eatTR{\begin{proof}
    It follows from Reflexivity (Lemma~\ref{lemma:reflexivity})
\end{proof}}
If $A \in X$, then $X \rightarrow A$ is a trivial dependency (Reflexivity).

\begin{lemma} (Optimization 2) \\
If  $X \rightarrow A$ is satisfied over $I$, then $XY \rightarrow A$ is satisfied for all $Y \subseteq R \setminus X$.
\vspace{-0.2cm}
\end{lemma}
\eatTR{\begin{proof}
    Assume $X \rightarrow A$. The OFD $X \rightarrow \{ \}$ follows from Reflexivity ((Lemma~\ref{lemma:reflexivity})). Hence, it can be inferred by Composition that $XY \rightarrow A$.
\end{proof}}
Intuitively, if $X \rightarrow A$ holds in $I$, then all OFDs containing supersets of $X$  also hold in $I$ (Augmentation), and can be pruned.  When we identify a key during OFD search, we can apply additional optimizations.

\begin{lemma} (Optimization 3) \\
If $X$ is a key (or super-key) in $I$, then for any attribute $A$,  $X \rightarrow A$ is satisfied in $I$.

\end{lemma}
\eatTR{\begin{proof}
    Since $X$ is a super-key, partition $\Pi_{X}$ consists of singleton equivalence classes only. Hence, the OFD $X \rightarrow A$ is valid.
\end{proof}}

For a candidate OFD $\set{X} \rightarrow \A{A}$, if $X$ is a key, then for all $x \in \Pi_{X}$, $|x|$ = 1, and $X \rightarrow A$ always holds.  On the other hand, if $\set{X} \setminus \A{A}$ is a superkey but not a key, then clearly the OFD $\set{X} \rightarrow \A{A}$ is not minimal. This is because there exists $\A{B} \in \set{X}$, such that $\set{X} \setminus \A{B}$ is a super key and $\set{X} \setminus \A{B} \rightarrow  \A{A}$ holds.

\begin{lemma} (Optimization 4) \\
If all tuples in an equivalence class $x \in \Pi_{X}$ have the same value of $A$, then a traditional FD, and therefore an OFD, is satisfied in $x$.

\end{lemma} 
\eatTR{\begin{proof}
Singleton equivalence classes over attribute set $X$ cannot falsify any OFD $X \rightarrow A$.
\end{proof}
}

A \emph{stripped partition} of $\Pi_{\set{X}}$, denoted   $\Pi^{*}_{\set{X}}$, removes all the equivalence classes of size one.  Single tuples cannot violate an OFD.
For example,
in Table~\ref{tab:mVio}, $\Pi_{\A{CC}}$ $=$ $\{ \brac{t_{1}, t_{5},t_{6}}, \brac{t_{2}, t_{4},t_{7}},$ $\brac{t_{3}} \}$, whereas the stripped partition removes the singleton equivalence class \{$t_{3}$\}, so $\Pi^{*}_{\A{CC}}$ $=$ $\{ \brac{t_{1}, t_{5},t_{6}}, \brac{t_{2}, t_{4},t_{7}} \}$.

\begin{lemma}\label{lemma:stripped}
Singleton equivalence classes over attribute set $\set{X}$ cannot violate any OFD $\set{X} \rightarrow \A{A}$.
\end{lemma}

\eatTR{
\begin{proof}
Follows directly from the definition of OFDs.  
\end{proof}
}

If $\Pi^{*}_{\set{X}}$ = $\emptySet$, then $\set{X}$ is a superkey and Optimization 3 (Lemma 3.6) applies.

\subsection{An Inference Procedure for OFDs}

Algorithm~\ref{alg:inference} computes the closure of a set of attributes given a set of OFDs. Our inference procedure can be applied to discovered, and subsequently user refined, OFDs to ensure continued minimality.  

\begin{algorithm}[t]
\begin{small}
\caption{Inference procedure for OFDs}\label{alg:inference}
\raggedright\textbf{Input}: A set of OFDs $\set{M}$, and a set of attributes
    $\set{X}$.\\
\textbf{Output}: The closure of $\set{X}$ with respect
    to $\set{M}$.

\begin{algorithmic}[1]
    \STATE $\set{M}_{unused}$ $\leftarrow$ $\set{M}$%
    \STATE $n$ $\leftarrow$ $0$
    \STATE $\set{X}^{n}$ $\leftarrow$ $\set{X}$
    \LOOP
    \IF{$\exists$ $\set{V}$ $\rightarrow$ $\set{Z}$ $\in$ \label{lines:st}
                    $\set{M}_{unused}$ and $\set{V}$ $\subseteq$ $\set{X}$}
            \STATE $\set{X}^{n+1}$ $\leftarrow$ $\set{X}^{n}$ $\cup$ $\set{Z}$
            \STATE $\set{M}_{unused}$ $\leftarrow$
            $\set{M}_{unused}$ $\setminus$ \{$\set{V}$ $\rightarrow$ $\set{Z}$\}
            \STATE $n$ $\leftarrow$ $n+1$ \label{lines:end}
        \ELSE
            \RETURN \emph{X}$^{n}$
        \ENDIF
    \ENDLOOP \STATE \textbf{end loop}

\end{algorithmic}
\end{small}
\end{algorithm}

\begin{theorem}
\label{thm:inference}
Algorithm~\ref{alg:inference} computes, in linear time, the closure $X^{+}$, $X^{+}$ $=$ $\{\A{A}$ $|$ $M$ $\vdash$ $X$ $\rightarrow$ $\A{A} \}$ , where $M$ denotes a set of OFDs.
\end{theorem}

\eatTR{\begin{proof}
First we show by induction on $k$ that if $\set{Z}$ is placed in $\set{X}^{k}$ in Algorithm \ref{alg:inference}, then $\set{Z}$ is in $\set{X}^{+}$.\\
\indent \emph{Basis}: $k$ = $0$. By Identity axiom $\set{X}$ $\rightarrow$ $\set{X}$.\\
\indent \emph{Induction}: $k$ $>$ $0$. Assume that $\set{X}^{k-1}$ consists only of attributes in $\set{X}^{+}$. Suppose $\set{Z}$ is placed in $\set{X}^{k}$ because $\set{V}$ $\rightarrow$ $\set{Z}$, and $\set{V}$ $\subseteq$ $\set{X}$. By Reflexivity $\set{X}$ $\rightarrow$ $\set{V}$, therefore, by Composition and Decomposition, $\set{X}$ $\rightarrow$ $\set{Z}$. Thus, $\set{Z}$ is in $\set{X}^{+}$.

Now we prove the opposite, if $\set{Z}$ is in $\set{X}^{+}$, then $\set{Z}$ is in the set returned by Algorithm \ref{alg:inference}. Suppose $\set{Z}$ is in $\set{X}^{+}$ but $\set{Z}$ is not in the set returned by Algorithm~\ref{alg:inference}.
Consider table $I$ similar to that in Table~\ref{table:tablet}. Table $I$ has three tuples that agree on attributes in $\set{X}$, are in a synonym or inheritance relationship, respectively, but not equal on \{$\set{X}^{n}$ $\setminus$ $\set{X}$\}, and are not in synonym nor inheritance relationship, respectively, on all other attributes (hence, also not equal).
We claim that $I$ satisfies $\set{M}$. If not, let $\set{P}$ $\rightarrow$ $\set{Q}$ be a dependency in $\set{M}$ that is violated by $I$. Then $\set{P}$ $\subseteq$ $\set{X}$ and $\set{Q}$ cannot be a subset of $\set{X}^{n}$, if the violation happens. Similar argument was
used in the proof of Theorem \ref{theorem:complete}. Thus, by Algorithm~\ref{alg:inference}, Lines~\ref{lines:st}--\ref{lines:end} there exists $\set{X}^{n+1}$, which is a contradiction.
\end{proof}}

\begin{example}
Let $\set{M}$ be the set of inheritance OFDs from Table~\ref{tab:example}: $\A{CC}$ $\rightarrow$ $\A{CTRY}$ and $\{ \A{CC, DIAG} \}$ $\rightarrow$ $\A{MED}$. Note that the inheritance OFD $\A{CC}$ $\rightarrow$ $\A{CTRY}$ holds since the synonym OFD $\A{CC}$ $\rightarrow_{syn}$ $\A{CTRY}$ holds and inheritance OFDs subsume synonym OFDs. The closure $\{ \A{CC, DIAG}\}^{+}$ computed with Algorithm~\ref{alg:inference} is $\{ \A{CC}$, \A{CTRY}, $\A{DIAG}$, $\A{MED} \}$.
\end{example}

For a given set of OFDs $\set{M}$, we can find an equivalent \emph{minimal} set, as defined below.

\begin{definition}
\label{def:cover}%
A set $\set{M}$ of OFDs is minimal if
\begin{enumerate}[nolistsep]
\item
    $\forall$ $\set{X} \rightarrow \set{Y} \in \set{M}$, $\set{Y}$ is a single attribute; \label{cond:one}
\item
     For no $\set{X}$ $\rightarrow$ $\A{A}$ and a proper subset $\set{Z}$ of $\set{X}$ is
    $\set{M}$ $\setminus$ $\{\set{X}$ $\rightarrow$ $\A{A}\}$ $\cup$ $\{\set{Z}$ $\rightarrow$ $\A{A}\}$
    equivalent to $\set{M}$; \label{cond:three}
\item
    For no $\set{X}$ $\rightarrow$ $\set{Y} \in \set{M}$ is $\set{M}$
    $\setminus$ $\{\set{X}$ $\rightarrow$ $\A{A}\}$ equivalent to $\set{M}$. \label{cond:two}
    \end{enumerate}
If $\set{M}$ is minimal and equivalent to a
set of OFDs $\set{N}$, then we say $\set{M}$ is a minimal cover of $\set{N}$.
\end{definition}

\begin{theorem}
\label{theorem:minimality}
Every set of OFDs $\set{M}$ has a minimal cover.
\end{theorem}

\eatTR{\begin{proof}
By the Union and Decomposition inference rules, it is possible to have $\set{M}$ with only a single attribute in the right hand side. We can achieve conditions two other conditions by repeatedly
deleting an attribute and then repeatedly removing a dependency. We can test whether
an attribute $\A{B}$ from $\set{X}$ is redundant for the OFD $\set{X}$ $\rightarrow$ $\A{A}$
by checking if $\A{A}$ is in $\{\set{X} \setminus \A{B}\}^{+}$. We can test whether $\set{X}$ $\rightarrow$ $\A{A}$ is redundant by computing closure $\set{X}^{+}$ with respect to $\set{M}$ $\setminus$ $\{\set{X}$
$\rightarrow$ $\A{A}\}$. Therefore, we eventually reach a set of OFDs which is
equivalent to $M$ and satisfies conditions \ref{cond:one},
\ref{cond:three} and \ref{cond:two}.
\end{proof}}

\begin{example}
Assume a set of inheritance OFDs $\set{M}$ = $\{ \set{M}_{1}: \A{CC}$ $\rightarrow$ $\A{CTRY}$, $\{ \set{M}_{2}: \A{CC}$, $\A{DIAG} \}$ $\rightarrow$ $\A{MED}$, $\{ \set{M}_{3}:$ $\A{CC}$, $\A{DIAG} \}$ $\rightarrow$ $\{ \A{MED}$, $\A{CTRY}\} \}$. This set is not a minimal cover as $\set{M}_{3}$ follows from $\set{M}_{1}$ and $\set{M}_{2}$ by Composition.
\end{example}


\section{Discovery Algorithm}
\label{sec:fastofd}

We now present an algorithm to discover a complete and minimal set of (synonym and inheritance) OFDs over a relation instance. Based on our axiomatization for OFDs (Section~\ref{sec:foundations}), we normalize all OFDs to a single attribute consequent, i.e., $X$ $\rightarrow$ $A$ for any attribute $A$.  An OFD $\set{X} \rightarrow \A{A}$ is \emph{trivial} if $\A{A} \in \set{X}$ by Reflexivity. An OFD $\set{X} \rightarrow \A{A}$ is \emph{minimal} if it is non-trivial and there is no set of attributes $\set{Y} \subset \set{X}$ such that $\set{Y} \rightarrow \A{A}$ holds in a table by Augmentation.

\begin{figure}[t]
\centering
       \includegraphics[width=2.6in]{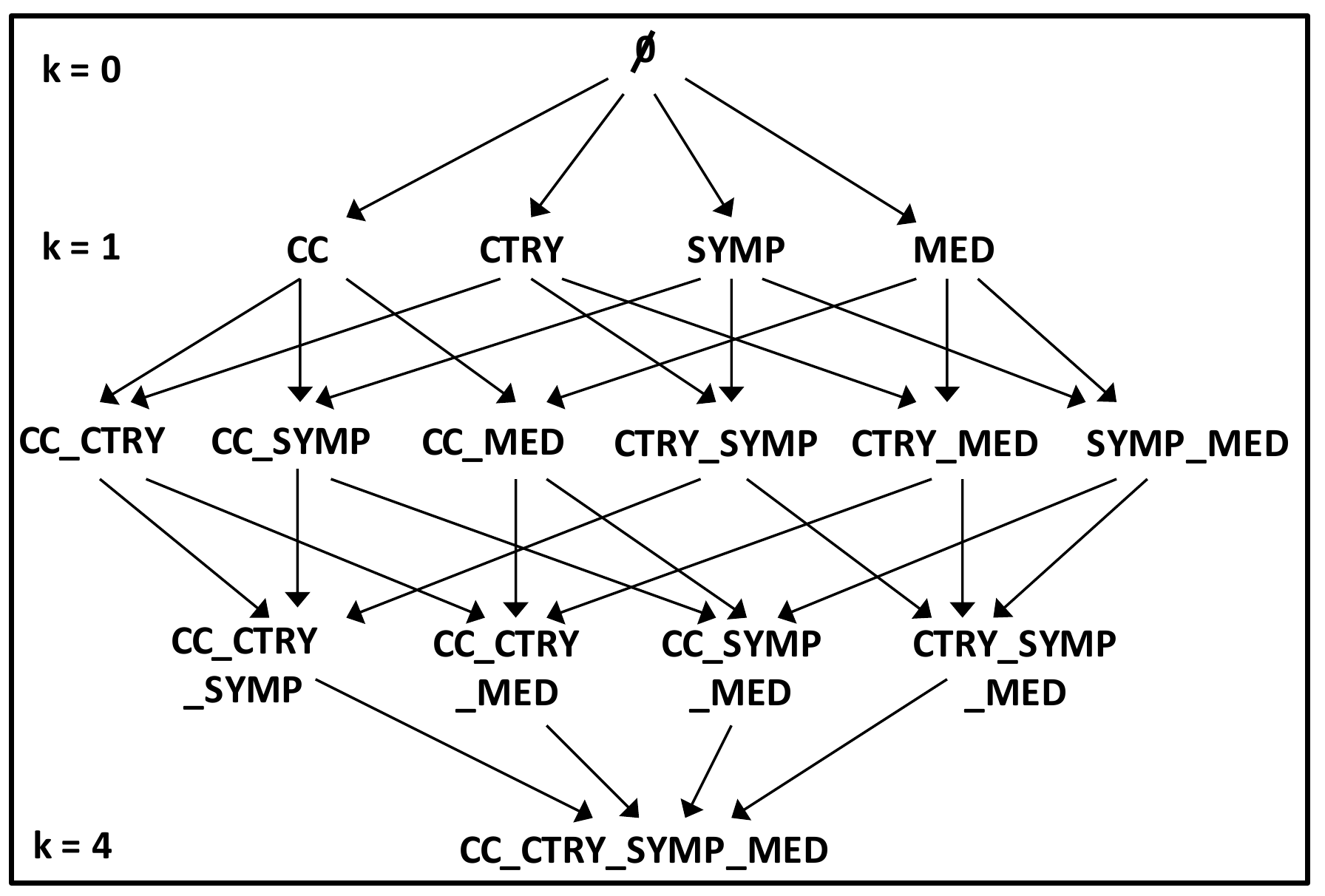}
         \vspace{-0.3cm}
    \caption{Example of an attribute lattice. \label{fig:lattice}}
    	\vspace{-0.3cm}
\end{figure}

The set of possible antecedent and consequent values considered by our algorithm can be modeled as a set containment lattice.
For example, Figure \ref{fig:lattice} shows the search lattice for four of the five attributes in Table \ref{tab:example}.  Each node in the lattice represents an attribute set and an edge exists between sets $X$ and $Y$ if $X$ $\subset$ $Y$ and $Y$ has exactly one more attribute than $X$.
Let $k$ be the number of levels in the lattice.  A relation with $n$ attributes will generate a $k = n$ level lattice, with $k = 0$ representing the top (root node) level.

After computing the stripped partitions $\Pi^*_{A}, \Pi^*_{B} \dots$ for single attributes at level $k = 1$, we efficiently compute the stripped partitions for subsequent levels in linear time by taking the product, i.e., $\Pi^*_{AB}$ = $\Pi^*_{A} \cdot \Pi^*_{B}$.  OFD candidates are considered by traversing the lattice in a breadth-first search manner.  We consider all $X$ consisting of single attribute sets, followed by all 2-attribute sets, and continue level by level to multi-attribute sets until (potentially) level $k$ = $n$, similarly as for other use cases of Apriori~\cite{AMW96}.  

\begin{algorithm}[t]
  \begin{small}
    \caption{FASTOFD} \label{pc:OFDdiscov}
\raggedright\textbf{Input:} Relation $\R{r}$ over schema $\R{R}$\\
   \textbf{Output:} Minimal set of OFDs $\set{M}$, such that $\R{r}$ $\models$ $\set{M}$

    \begin{algorithmic}[1]
        \STATE $\set{L}_{0}$ $=$ $\emptySet{}$
        \STATE $\set{C}^{+}(\emptySet{}) = \R{R}$
        \STATE $l$ $=$ $1$
        \STATE $\set{L}_{1}$ $=$ $\{ \A{A}$ $|$ $\A{A} \in \R{R} \}$
        \WHILE{$\set{L}_{l} \not= \emptySet$} \label{line:mainLoop}
            \STATE \emph{computeOFDs}($\set{L}_{l}$)
            \STATE $\set{L}_{l+1}$ $=$ \emph{calculateNextLevel}($\set{L}_{l}$)
            \STATE $l$ $=$ $l+1$
        \ENDWHILE \label{line:endMainLoop}
        \RETURN $\set{M}$
    \end{algorithmic}
  \end{small}
\end{algorithm}
\begin{table}\vskip -0.75cm \end{table}

Algorithm~\ref{pc:OFDdiscov} shows the skeleton of our OFD discovery process.
In level  $\set{L}_{l}$ of the search lattice, our algorithm generates candidate OFDs with $l$ attributes using  \emph{computeOFDs}($\set{L}_{l})$. \emph{FASTOFD} starts the search from singleton sets of attributes and works its way to larger attribute sets through the lattice, level by level. When the algorithm processes an attribute set $\set{X}$, it verifies candidate OFDs of the form ($\set{X} \setminus \A{A}) \rightarrow \A{A}$, where $\A{A} \in \set{X}$.  This guarantees that only non-trivial OFDs are considered.  For each candidate, we check if it is a valid synonym or inheritance OFD as per Definition~\ref{def:synonym_ofd} and Definition~\ref{def:inheritance_ofd}.

The small-to-large search strategy guarantees that only OFDs that are minimal are added to the output set of OFDs $\set{M}$, and is used to prune the search space effectively. The OFD candidates generated in a given level are checked for \emph{minimality} based on the previous levels and are added to a \emph{valid} set of OFDs $\set{M}$ if applicable. The algorithm \emph{calculateNextLevel}($\set{L}_{l}$) forms the next level from the current level.

Next, we explain, in turn, each of the algorithms that are called in the main loop of \emph{FASTOFD}.

\subsection{Finding Minimal OFDs}\label{sec:optimizations}

\emph{FASTOFD} traverses the lattice until all minimal OFDs are found.
We deal with OFDs of the form $\set{X} \setminus \A{A} \rightarrow \A{A}$, where $\A{A} \in \set{X}$.  To check if such an OFD is minimal, we need to know if $\set{X} \setminus \A{A} \rightarrow \A{A}$ is valid for $\set{Y} \subset \set{X}$. If $\set{Y} \setminus \A{A} \rightarrow \A{A}$, then by Augmentation $\set{X} \setminus \A{A} \rightarrow \A{A}$ holds. An OFD $\set{X} \rightarrow \A{A}$ holds for any relational instance by Reflexivity, therefore, considering only $\set{X} \setminus \A{A} \rightarrow \A{A}$ guarantees that only non-trivial OFDs are taken into account.

We maintain information about minimal OFDs, in the form of $\set{X} \setminus \A{A} \rightarrow \A{A}$, in the \emph{candidate} set $\set{C}^{+}(\set{X})$. If $\A{A} \in \set{C}^{+}(\set{X})$ for a given set $\set{X}$, then $\A{A}$ has not been found to depend on any proper subset of $\set{X}$. Therefore, to find minimal OFDs, it suffices to verify OFDs $\set{X} \setminus \A{A} \rightarrow \A{A}$, where $\A{A} \in \set{X}$ and $\A{A} \in \set{C}^{+}(\set{X} \setminus \A{B})$ for all $\A{B} \in \set{X}$.

\begin{example}
Assume that $\A{B} \rightarrow \A{A}$ and that we consider the set $\set{X}$ = $\brac{\A{A}, \A{B}, \A{C}}$. As $\A{B} \rightarrow \A{A}$ holds, $\A{A} \not \in$ $\set{C}^{+}(\set{X} \setminus \A{C})$. Hence, the OFD $\brac{\A{B}, \A{C}} \rightarrow \A{A}$ is not minimal.
\end{example}

We define the candidate set $\set{C}^{+}(\set{X})$ as follows.%
\footnote{Some of our techniques are similar to \emph{TANE}~\cite{HKPT98} for FD discovery and \emph{FASTOD}~\cite{ODTR16} for Order Dependency (OD) discovery since OFDs subsume FDs and ODs subsume FDs. However, FASTOFD differs in many details from \emph{TANE} and \emph{FASTOD}, e.g., optimizations, removing nodes from the lattice and key pruning rules.  Also, \emph{FASTOFD} includes OFD-specific rules. For instance, for FDs if $\brac{\A{B}, \A{C}} \rightarrow \A{A}$ and $\A{B} \rightarrow \A{C}$, then $\A{B} \rightarrow \A{A}$ holds, hence, $\brac{\A{B}, \A{C}} \rightarrow \A{A}$ is considered non-minimal. However, this rule does not hold for OFDs, and therefore our definition of candidate set $\set{C}^{+}(\set{X})$ differs from \emph{TANE}.}

\begin{definition}\label{def:minConst}
$\set{C}^{+}(\set{X})$ $=$ $\{ \A{A} \in \R{R}$ $|$ \\
     \indent \indent \indent $\forall_{\A{A} \in \set{X}}$  $\set{X} \setminus \A{A} \rightarrow \A{A}$ does not hold$\}$.
\end{definition}

\subsection{Computing Levels}\label{sec:calcLev}

Algorithm~\ref{pc:nextLevel} explains \emph{calculateNextLevel}($\set{L}_{l}$), which computes $\set{L}_{l+1}$ from $\set{L}_{l}$. It uses the subroutine \emph{singleAttrDifferBlocks}($\set{L}_{l}$) that partitions $\set{L}_{l}$ into blocks (Line~\ref{line:blocks}). Two sets belong to the same block if they have a common subset $\set{Y}$ of length $l-1$ and differ in only one attribute, $\A{A}$ and $\A{B}$, respectively. Therefore, the blocks are not difficult to calculate as sets $\set{Y}\A{A}$ and $\set{Y}\A{B}$ can be expressed as sorted sets of attributes. Other usual use cases of Apriori~\cite{AMW96} such as \emph{TANE}~\cite{HKPT98}
use a similar approach.

\begin{algorithm}[t]
  \begin{small}
    \caption{calculateNextLevel($\mathcal{L}_{l}$)} \label{pc:nextLevel}

    \begin{algorithmic}[1]
        \STATE $\set{L}_{l+1} = \emptySet{}$
        \FORALL{$\brac{\set{Y}\A{B}, \set{Y}\A{C}} \in \emph{singleAttrDiffBlocks}(\set{L}_{l})$} \label{line:blocks}
            \STATE $\set{X} = \set{Y} \bigcup \brac{\A{B}, \A{C}}$
            \label{line:prevLevel}
            \STATE Add $\set{X}$ to $\set{L}_{l+1}$
        \ENDFOR
        \RETURN $\set{L}_{l+1}$
    \end{algorithmic}
  \end{small}
\end{algorithm}

The level  $\set{L}_{l+1}$ contains those sets of attributes of size $l + 1$ which have their subsets of size $l$ in $\set{L}_{l}$.

\subsection{Computing Dependencies \& Completeness}

Algorithm~\ref{pc:computeODs}, \emph{computeOFDs}($\set{L}_{l}$), adds minimal OFDs from level $\set{L}_l$ to $\set{M}$, in the form of $\set{X} \setminus \A{A} \rightarrow \A{A}$, where $\A{A} \in \set{X}$. The following lemma shows that we can use the candidate set $\set{C}^{+}(\set{X})$ to test whether  $\set{X} \setminus \A{A} \rightarrow \A{A}$ is minimal.

\begin{lemma}\label{minConst}
An OFD $\set{X} \setminus \A{A} \rightarrow \A{A}$, where $\A{A} \in \set{X}$, is minimal iff $\forall_{\A{B} \in \set{X}} \A{A} \in \set{C}^{+}(\set{X} \setminus \A{B})$.
\end{lemma}

\eatTR{
\begin{proof}
Assume first that the dependency $\set{X} \setminus \A{A} \rightarrow \A{A}$ is not minimal. Therefore, there exists $\A{B} \in \set{X}$ for which $\set{X} \setminus \brac{\A{A}, \A{B}} \rightarrow \A{A}$ holds. Then, $\A{A} \not \in \set{C}^{+}(\set{X} \setminus \A{B})$.

To prove the other direction assume that there exists $\A{B} \in \set{X}$, such that $\A{A} \not \in \set{C}^{+}(\set{X} \setminus \A{B})$. Therefore, $\set{X} \setminus \brac{\A{A}, \A{B}} \rightarrow \A{A}$ holds, where $\A{A} \not = \A{B}$. Hence, by Reflexivity the dependency $\set{X} \setminus \A{A} \rightarrow \A{A}$ is not minimal.
\end{proof}
}

\begin{algorithm}[t]
  \begin{small}
    \caption{computeOFDs($\mathcal{L}_{l}$)} \label{pc:computeODs}

    \begin{algorithmic}[1]
        \FORALL{$\set{X} \in \set{L}_{l}$} \label{line:loopLevel}
                \STATE $\set{C}^{+}(\set{X}) = \bigcap_{\A{A} \in \set{X}} \set{C}^{+}_{c}(\set{X} \setminus \A{A})$ \label{line:interConst}
        \ENDFOR

        \FORALL{$\set{X} \in \set{L}_{l}$}
            \FORALL{$\A{A} \in \set{X} \cap \set{C}^{+}(\set{X})$}
                    \label{line:candConst}
                \IF{$\set{X} \setminus \A{A} \rightarrow \A{A}$}
                        \label{line:checkConst}
                    \STATE Add $\set{X} \setminus \A{A} \rightarrow \A{A}$
                            to $\set{M}$ \label{line:addConst}
                    \STATE Remove $\A{A}$ from $\set{C}^{+}(\set{X})$
                    \label{line:removeAttr}
                \ENDIF
            \ENDFOR
        \ENDFOR
    \end{algorithmic}
  \end{small}
\end{algorithm}

By Lemma~\ref{minConst}, the steps in Lines~\ref{line:interConst}, \ref{line:candConst}, \ref{line:checkConst} and \ref{line:addConst} guarantee that the algorithm adds to $\set{M}$ only the minimal OFDs of the form $\set{X} \setminus \A{A} \rightarrow \A{A}$, where $\set{X} \in \set{L}_{l}$ and $\A{A} \in \set{X}$.   In Line \ref{line:checkConst}, to verify whether  $\set{X} \setminus \A{A} \rightarrow \A{A}$ is a synonym or inheritance OFD, we apply Definition~\ref{def:synonym_ofd} and Definition~\ref{def:inheritance_ofd}, respectively.

\begin{lemma}\label{lemma:levelCorrSplit}
Let candidates $\set{C}^{+}(\set{Y})$ be correctly computed $\forall \set{Y} \in \set{L}_{l-1}$.  $computeOFDs$($\set{L}_{l}$) calculates correctly $\set{C}^{+}(\set{X})$, $\forall \set{X} \in \set{L}_{l}$.
\end{lemma}

\eatTR{
\begin{proof}
An attribute $\A{A}$ is in $\set{C}^{+}(\set{X})$ after the execution of the algorithm $computeOFDs$($\set{L}_{l})$  unless it is excluded from $\set{C}^{+}(\set{X})$ on Line \ref{line:interConst} or \ref{line:removeAttr}. First we show that if $\A{A}$ is excluded from $\set{C}^{+}(\set{X})$ by $computeOFDs$($\set{L}_{l})$, then $\A{A} \not \in$ $\set{C}^{+}(\set{X})$ by the definition of $\set{C}^{+}(\set{X})$.
    \begin{itemize}[nolistsep]
        \item[-] If $\A{A}$ is excluded from $\set{C}^{+}(\set{X})$ on Line~\ref{line:interConst}, there exists $\A{B} \in \set{X}$ with $\A{A} \not \in \set{C}^{+}(\set{X} \setminus \A{B})$. Therefore, $\set{X} \setminus \brac{\A{A}, \A{B}} \rightarrow \A{A}$ holds, where $\A{A} \not = \A{B}$. Hence, $\A{A} \not \in$ $\set{C}^{+}(\set{X})$ by the definition of $\set{C}^{+}(\set{X})$.
        \item[-] If $\A{A}$ is excluded on Line~\ref{line:removeAttr}, then $\A{A} \in \set{X}$ and $\set{X} \setminus \A{A} \rightarrow \A{A}$ holds. Hence, $\A{A} \not \in$ $\set{C}^{+}(\set{X})$ by the definition of $\set{C}^{+}(\set{X})$.
    \end{itemize}
Next, we show the other direction, that if $\A{A} \not \in$ $\set{C}^{+}(\set{X})$ by the definition of $\set{C}^{+}(\set{X})$, then $\A{A}$ is excluded from $\set{C}^{+}(\set{X})$ by the algorithm $computeOFDs$($\set{L}_{l}$). Assume $\A{A} \not \in$ $\set{C}^{+}(\set{X})$ by the definition of $\set{C}^{+}(\set{X})$. Therefore, there exists $\A{B} \in \set{X}$, such that $\set{X} \setminus \brac{\A{A}, \A{B}} \rightarrow \A{A}$ holds. We have following two cases.
    \begin{itemize}[nolistsep]
        \item[-] $\A{A} = \A{B}$. Thus, $\set{X} \setminus \A{A} \rightarrow \A{A}$ holds and $\A{A}$ is removed on Line~\ref{line:removeAttr}, if $\set{X} \setminus \A{A} \rightarrow \A{A}$ is minimal; and on Line~\ref{line:interConst} otherwise.
        \item[-] $\A{A} \not = \A{B}$. Hence, $\A{A} \not \in$ $\set{C}^{+}(\set{X} \setminus \A{B})$ and $\A{A}$ is removed on Line~\ref{line:interConst}.
    \end{itemize}
This ends the proof of correctness of computing the candidate set $\set{C}^{+}(\set{X})$, $\forall \set{X} \in \set{L}_{l}$.
\end{proof}
}

Next, we show that our OFD discovery algorithm produces a complete and minimal set of OFDs.

\begin{theorem}\label{theorem:completeness}
The $FASTOFD$ algorithm computes a complete and minimal set of OFDs $\set{M}$.
\end{theorem}

\eatTR{
\begin{proof}
The algorithm $computeOFDs$($\set{L}_{l})$ adds to set of OFDs $\set{M}$ only the minimal OFDs. The steps in Lines~\ref{line:interConst}, \ref{line:candConst}, \ref{line:checkConst} and \ref{line:addConst} guarantee that the algorithm adds to $\set{M}$ only the minimal OFDs of the form $\set{X} \setminus \A{A} \rightarrow \A{A}$, where $\set{X} \in \set{L}_{l}$ and $\A{A} \in \set{X}$ by Lemma~\ref{minConst}. It follows by induction that $computeOFDs$($\set{L}_{l})$ calculates correctly $\set{C}^{+}(\set{X})$ for all levels $l$ of the lattice since Lemma~\ref{lemma:levelCorrSplit} holds. Therefore, the FASTOFD algorithm computes a complete set of minimal OFDs $\set{M}$.
\end{proof}
}


\subsection{Complexity Analysis}
\label{sec:complexity}
The algorithm complexity depends on the number of candidates in the lattice.
The worst case complexity of our algorithm is \emph{exponential} in the number of attributes as there are $2^{|n|}$ nodes in the search lattice.  Furthermore, the worst-case output size is also exponential in the number of attributes (and occurs when the minimal OFDs are in the widest middle level of the lattice).  This means that a polynomial-time discovery algorithm in the number of attributes cannot exist.  These results are in line with previous FD \cite{HKPT98}, inclusion dependency \cite{PKQJN15}, and order dependency~\cite{ODTR16} discovery algorithms.


However, the complexity is polynomial in the number of tuples, although the ontological relationships (synonyms and inheritance) influence the complexity of verifying whether a candidate OFD holds. We assume that values in the ontology are indexed and can be accessed in constant time.  To verify whether a candidate synonym OFD holds over $I$, for each equivalence class $x \in \Pi_{X}(I)$, we need to check whether the intersection of the corresponding senses is not empty (Definition~\ref{def:synonym_ofd}).  This can be done in linear time (in the number of tuples) by scanning the stripped partitions and maintaining a hash table with the frequency counts of all the senses for each equivalence class.  Returning to the example in Table~2, the synonym OFD $X \rightarrow_{syn} Y$ does not hold because for the single equivalence class in this example, of size three, there are no senses (classes) for $Y$ that appear three times.  For inheritance OFDs, the complexity grows to cubic in the number of tuples because of the additional task of locating the LCA for each equivalence class, which takes quadratic time.
\subsection{Approximate OFDs}
\label{sec:approx}
Up to now, we have focused on the discovery of OFDs that hold over the entire relational instance $I$.
In practice, some applications do not require such a strict notion of satisfaction, and OFDs may not hold exactly over the entire relation due to errors in the data.
In such cases, approximate OFDs, which hold over a subset of $I$, are more appropriate.  Similar to previous work on approximate FD discovery, we define a minimum support level, $\tau$, that defines the minimum number of tuples that must satisfy an OFD $\phi$.   We define the problem of approximate OFD discovery as follows.  Given a relational instance $I$, and a minimum support threshold $\tau, 0 \leq \tau \leq 1$, we want to find all minimal OFDs $\phi$ such that $s(\phi) \geq \tau$ where $s(\phi)$ =  $max$ \{$|r| \mid $  $r \subseteq I$, $r \models \phi$\}.

The main modification to discover approximate OFDs is in the verification step of checking whether a candidate is a synonym or an inheritance OFD.  The candidate generation and optimization steps remain the same.  This requires first identifying the tuples participating in a synonymous or inheritance relationship, and checking whether the number of satisfying tuples is greater than or equal to $\tau$.  For synonyms, we check for the maximum number of values that resolve to the same sense in each equivalence class;
then, we check whether the number of satisfying tuples exceeds our minimum support level $\tau$.  By a similar argument to that in Section~\ref{sec:complexity} for exact synonym OFDs, the complexity remains linear in the number of tuples.  The difference here is that we use the hash table storing frequency counts of different senses to find the \emph{most frequently occurring senses} for each equivalence class.  The sum of these counts gives us the number of tuples satisfying the OFD.  A similar argument holds for approximate inheritance OFDs.

In Section \ref{sec:ofdclean}, we experimentally verify the usefulness of approximate OFDs in data cleaning.  Given that approximate OFDs hold over a subset of the relation, if we remove or correct a number of ``dirty" tuples, then we expect the OFD to be satisfied over the entire relation.  We evaluate the number of approximate OFDs found for varying support levels $\tau$.  We  show that approximate OFDs are a useful dependency to not only identify potentially dirty data values, but based on the approximate OFD, we can recommend possible repairs to correct these inconsistent values.


\section{Experiments}
\label{sec:eval}

We now turn to the evaluation of our techniques using real datasets.  Our evaluation focuses on four objectives:

\begin{enumerate}[nolistsep,leftmargin=*]
\item An evaluation of the efficiency of our algorithm to identify interesting dependencies early on during the lattice-based traversal.  We measure efficiency in terms of running time and the size of the OFDs discovered at each level of the lattice.
\item An evaluation of the scalability and performance as compared to existing FD discovery algorithms.  We evaluate our algorithm performance against seven existing techniques as we scale the number of tuples and the number of attributes. 
\item An evaluation of the performance benefits of our proposed optimization techniques to prune redundant OFD candidates.
\item A qualitative evaluation of the utility of the discovered OFDs.
\end{enumerate}

Our experiments were performed using four Intel Xeon processors at 2.1GHz each with 32 GB of memory.  All algorithms were implemented in Java.  The reported runtimes are averaged over six executions.  For  the comparative experiments, we use the implementations for FD discovery provided by the Metanome profiling platform \cite{PBF+15}. 

\subsection{Data Characteristics}
We use two real datasets.  The first dataset is obtained from the Linked Clinical Trials (LinkedCT.org) database.  The LinkedCT.org project provides an open interface for international clinical trials data. We use 1M records (15 attributes) including data about the study, country, medical diagnosis, prescribed drugs, illnesses, symptoms, treatment, and outcomes.  The second dataset contains US census data (collected from http://census.gov/).  The data contains population properties such as work class, marital status, race, relationship, occupation, and salary.  We used a portion of this dataset that contains a total of 150K records with 11 attributes.  In all our experiments, we refer to the U.S. National Library of Medicine Research \cite{medOntology}, and WordNet ontologies.


\subsection{Scalability}
\begin{figure*}[ht]
\centering

\minipage[b]{0.3\linewidth}
  \centering
  \includegraphics[width=1.9in]{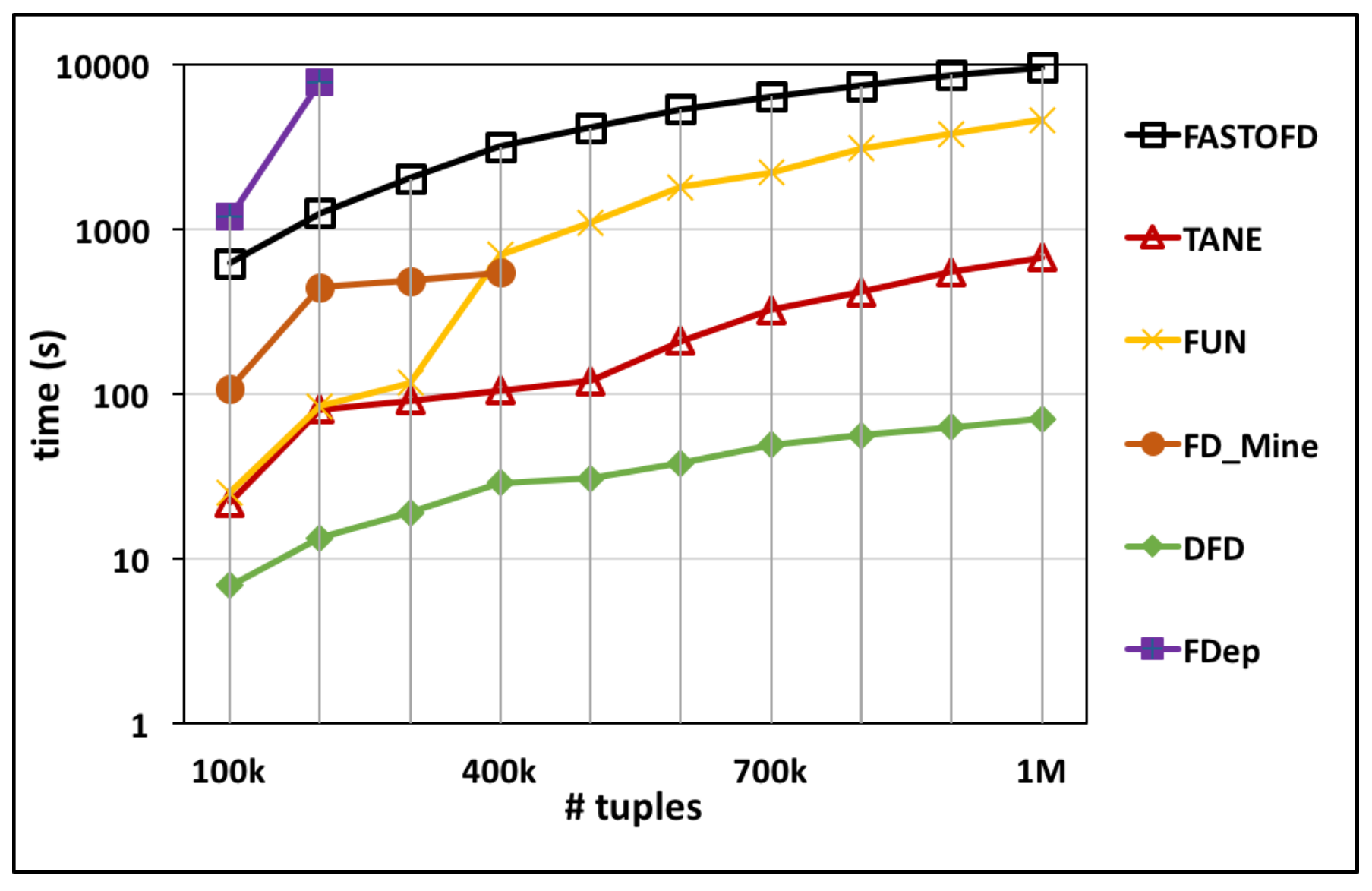}
  \vspace{-0.3cm}
  \caption{Scalability in N (clinical).}\label{fig:clinical}
\endminipage
\minipage[b]{0.3\linewidth}
   \centering
  \includegraphics[width=1.9in]{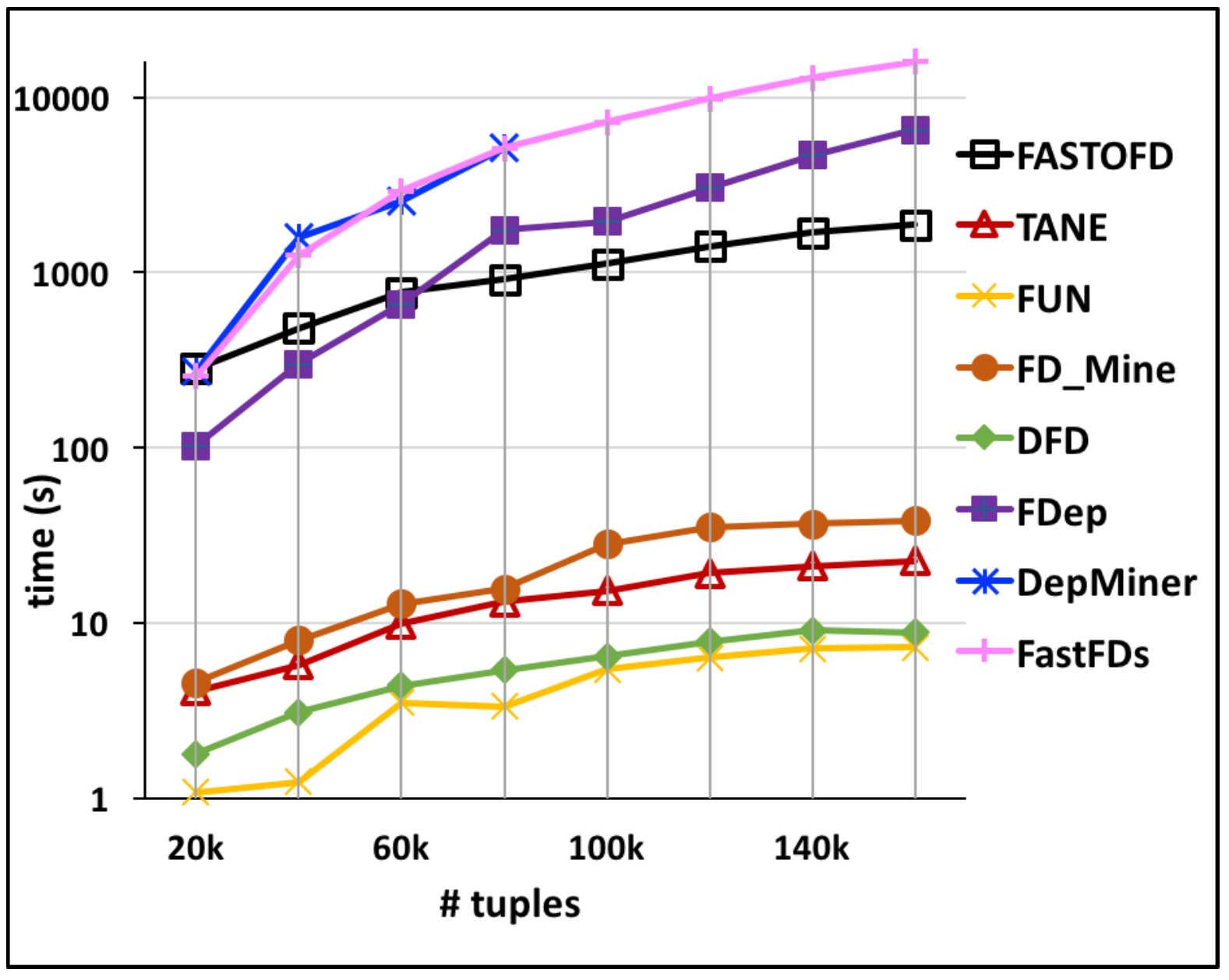}
  \vspace{-0.3cm}
\caption{Scalability in N (census). \label{fig:census}}
\endminipage
\minipage[b]{0.3\linewidth}
   \centering
 \includegraphics[width=1.9in]{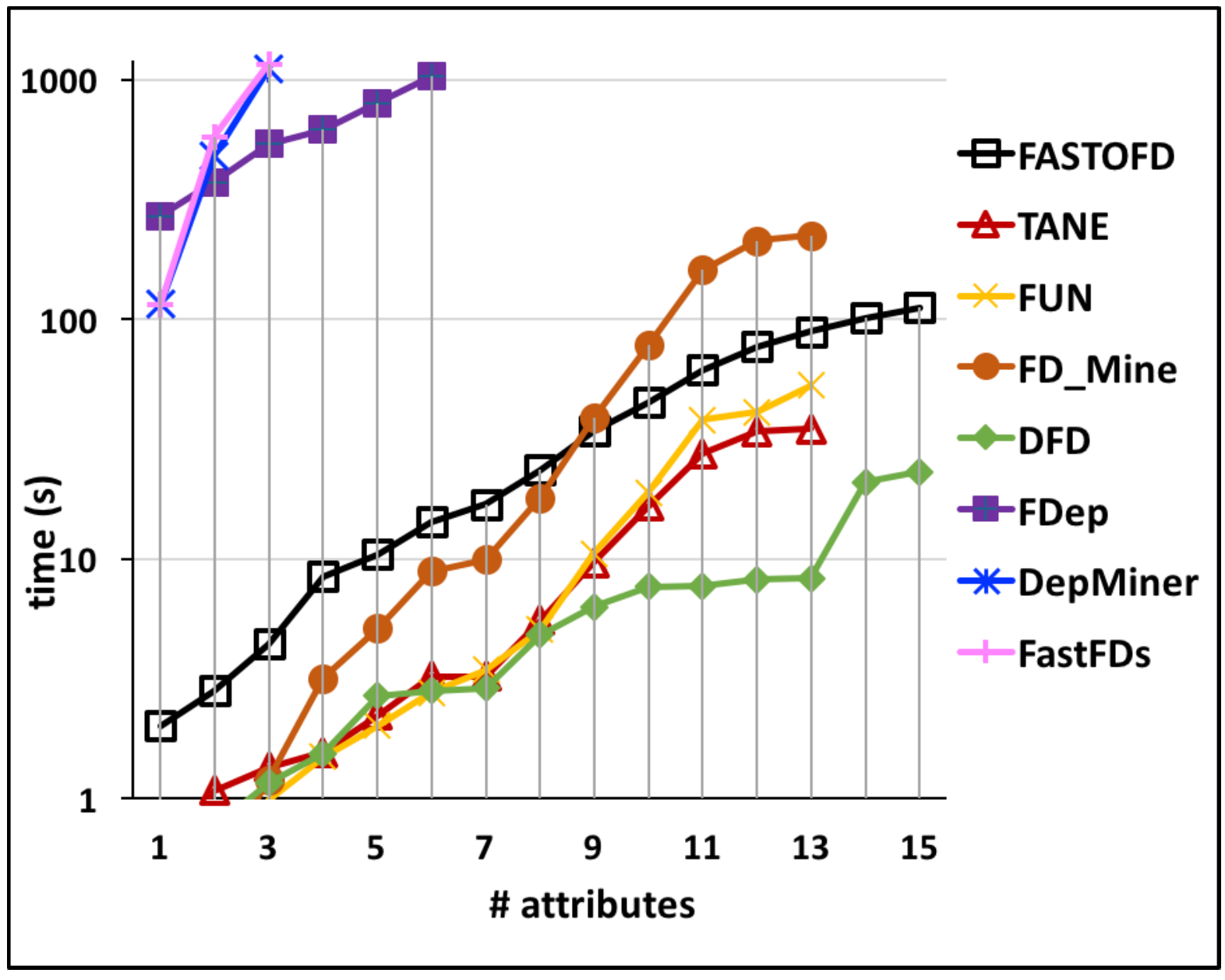}
 \vspace{-0.3cm}
  \caption{Scalability in $n$ (clinical).}\label{fig:numattrs}
   \endminipage\hfill
   \minipage[b]{0.3\linewidth}
  \centering
 \includegraphics[width=1.9in]{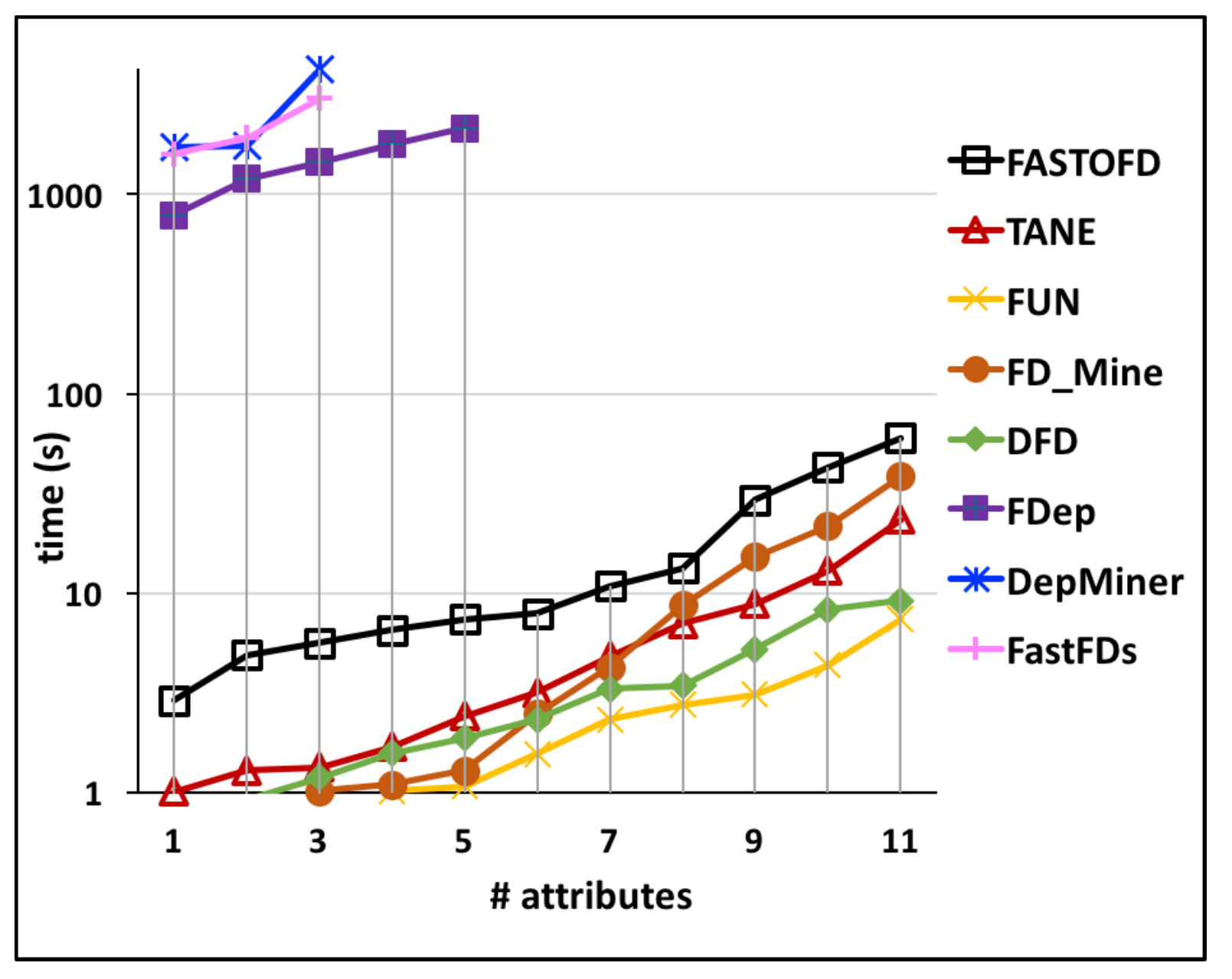}
 \vspace{-0.3cm}
  \caption{Scalability in $n$ (census).}\label{fig:numattrs2}
 \vspace{-0.3cm}
\endminipage
\minipage[b]{0.3\linewidth}
   \centering
  \includegraphics[width=1.9in]{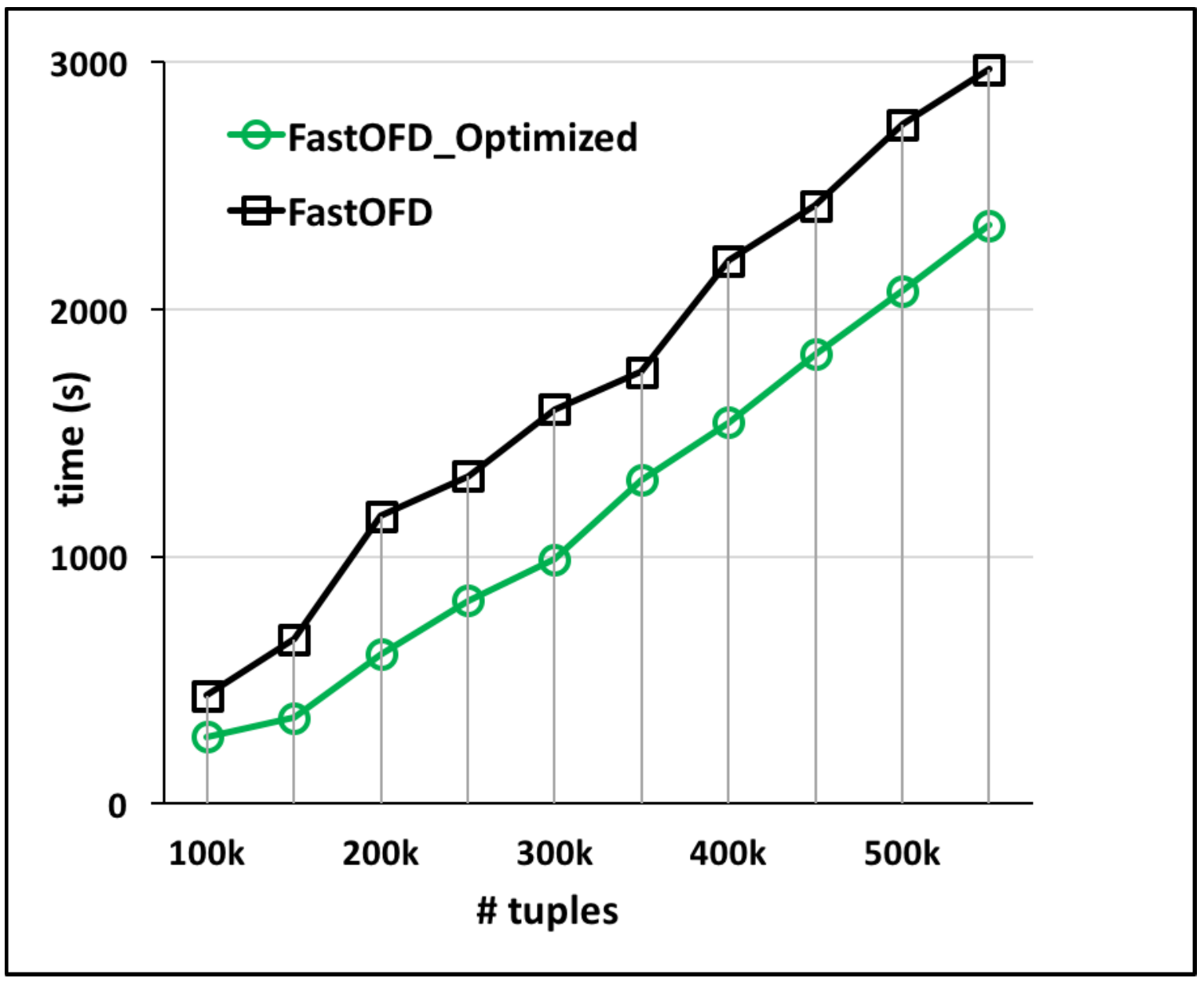}
 \vspace{-0.3cm}
  \caption{Impact of Opt-2.}\label{fig:opt2}
  \vspace{-0.3cm}
\endminipage
\minipage[b]{0.3\linewidth}
   \centering
 \includegraphics[width=1.9in]{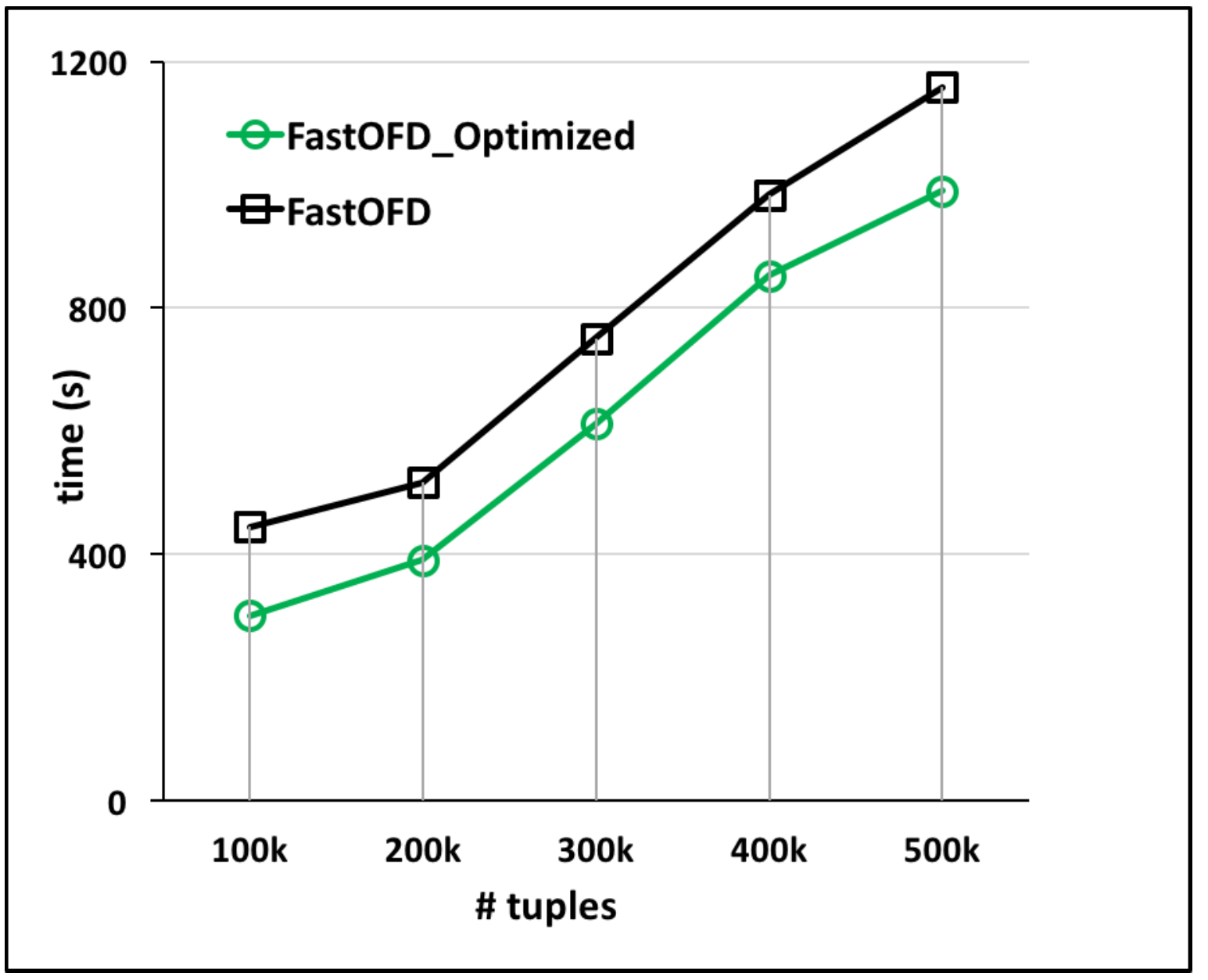}
 \vspace{-0.3cm}
  \caption{Impact of Opt-3.}\label{fig:opt3}
  \vspace{-0.3cm}
   \endminipage\hfill
   
\end{figure*}

\textbf{Experiment-1: Comparative Scalability in Number of Tuples.}  We evaluate the scalability of our algorithm with respect to the number of tuples ($N$) against seven existing FD discovery algorithms: TANE \cite{HKPT98}, FUN \cite{NC01}, FDMine \cite{YH08}, DFD \cite{ASN14}, DepMiner \cite{LPL00}, FastFDs \cite{WGR01}, and FDep \cite{FS99}.  Figure \ref{fig:clinical} and Figure \ref{fig:census} show the running times using the clinical trials (with 15 columns) and census (11 columns) datasets, respectively, using $\theta = 5$.  In Figure \ref{fig:clinical}, we report partial results for FDMine and FDep as both techniques exceeded the main memory limits.  Similar to previous studies, we found that FDMine returns a much larger number of non-minimal dependencies, about 24x (for clinical) and 118x (for census) leading to increased memory requirements \cite{ASN14}.   We ran DepMiner and FastFDs using 100K records and report running times of 4hrs and 2.3 hrs, respectively. However, for larger data sizes (200K+ records), we terminated runs for these two techniques after 12 hours, hence they do not appear in Figure \ref{fig:clinical}.  
For the smaller census data, Figure \ref{fig:census} shows that our techniques outperform  FastFDs, DepMiner, and FDep, since these three techniques scale quadratically with respect to the number of tuples.

The running times for FASTOFD scale linearly with the number of tuples, similar to other lattice traversal based approaches (TANE, FUN, and DFD). The runtime is dominated by data verification of OFDs. 
As expected, discovering OFDs incurs an increased runtime.  Specifically, we found that discovering synonym and inheritance OFDs incur an average overhead of 1.8x and 2.4x, respectively, over existing lattice-traversal FD discovery algorithms.  This can be explained by the inherent complexity of OFDs (which subsume FDs), and the increased number of discovered OFDs. For example, in the clinical data, many illnesses and medications are referenced by multiple names but refer to the same entity; e.g., a drug can be referred to by its generic name or its brand name.  Given the increased number and increased complexity of OFDs, FASTOFD discovers these dependencies within reasonable time bounds from existing solutions.

\eatTR{
Similar to previous studies, we found that FD
Our techniques show a linear scale-up w.r.t. the number of tuples, with the majority of time dominated by verification of OFDs.  For our chosen datasets, we found there was a large number of smaller equivalence classes, which lead to the decreased verification time that dominated the overall running time.  Our techniques outperform TANE by an average of 19.5\%, and we observed that our clinical trials dataset contained 3x more satisfying OFDs versus FDs, since many illnesses and medications are referenced by multiple names but refer to the same entity, e.g., a drug can be referred to by its generic name or its brand name.   We observe that discovering inheritance FDs incurs an average of 7.1\% overhead than discovering synonym FDs.  For inheritance FDs, we need to traverse the ontology to find the least common ancestor within a path length of $\theta$.  In contrast, the verification of a synonym FDs requires checking for a non-empty intersection among a set of values, which can be done in constant time.
}

\textbf{Experiment-2: Comparative Scalability in Number of Attributes.}   We evaluate FASTOFD's scalability as we scale the number of attributes ($n$), using $N$ = 100k tuples, and $\theta = 5$.  Figure \ref{fig:numattrs} and Figure \ref{fig:numattrs2} show  the running times using the clinical trials and census datasets, respectively.  We observe that all algorithms scale exponentially w.r.t. the number of attributes since the space of candidates grows with the number of attribute set combinations.  FASTOFD scales comparatively with other lattice based approaches.  Our solution discovers 3.1x more dependencies on average (this includes synonym OFDs, inheritance OFDs, and traditional FDs since they are subsumed by OFDs), compared to existing approaches, validating the overhead we incur.  In Figure \ref{fig:numattrs} and Figure \ref{fig:numattrs2}, we report partial results for DepMiner, FastFDs, and FDep (before memory limits were exceeded), where we achieve almost two orders of magnitude improvement due to our optimizations.  Similar to existing lattice-based approaches, our techniques performs well on a smaller number of attributes due to effective pruning strategies that reduce the search space.  

\eatTR{
As expected, we saw an exponential scale-up in running time w.r.t. the number of attributes.  This is not surprising given that the number of minimal OFDs over the attribute lattice is exponential in the worst case. 
A larger number of inheritance versus synonym FDs are found, leading to the increased running time.  We observe that the execution time more than triples from $n = 7$ to $n = 15$ for clinical data, and more than 5x from $n = 7$ to $n = 11$ for census data, due to the increased number of attribute combinations, and consequently, the larger search space of candidates that must be evaluated.   
}

\begin{figure*}[ht]
\centering
   \minipage[b]{0.3\linewidth}
   \centering
 \includegraphics[width=1.9in]{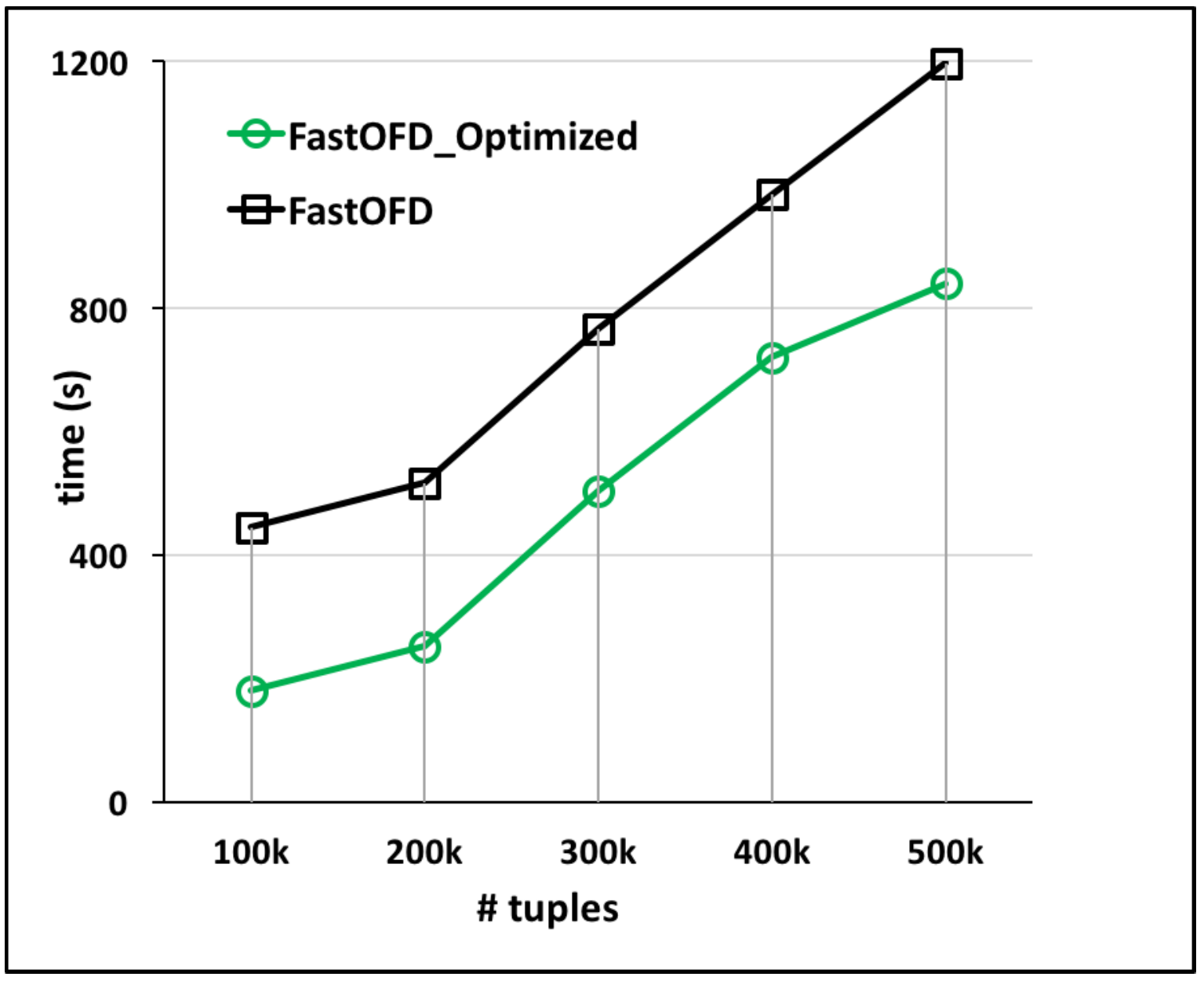}
 \vspace{-0.3cm}
  \caption{Impact of Opt-4.}\label{fig:opt4}
  \vspace{-0.1cm}
   \endminipage
   \minipage[b]{0.3\linewidth}
   \centering
  \includegraphics[width=1.6in]{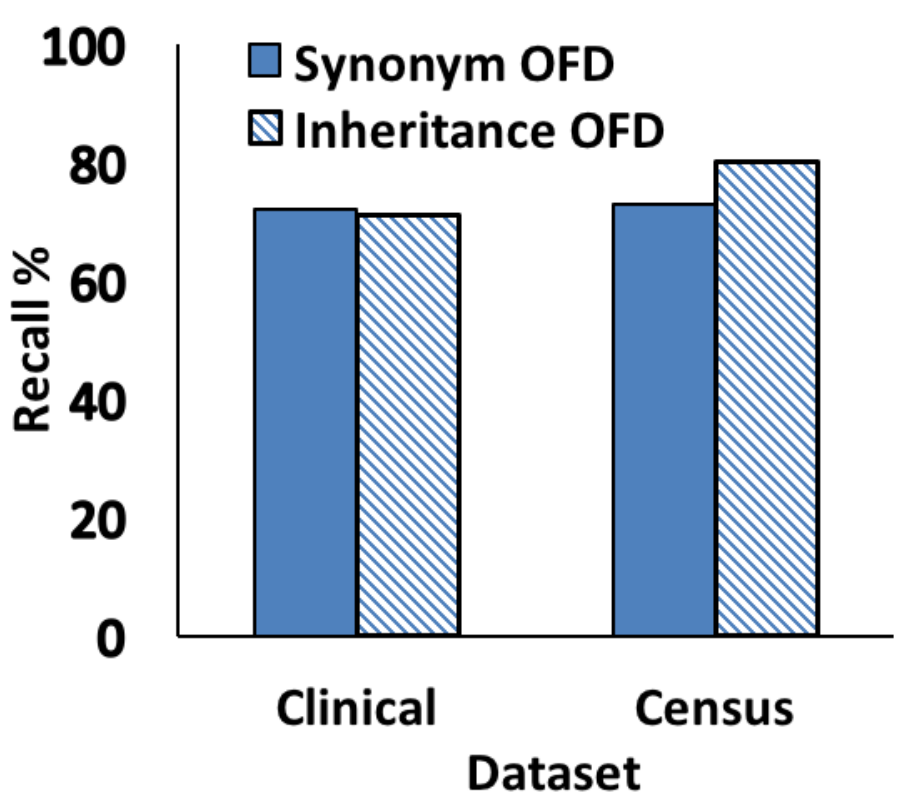}
 \vspace{-0.3cm}
  \caption{Recall.}\label{fig:recall}
  \vspace{-0.1cm}
   \endminipage
   \minipage[b]{0.3\linewidth}
   \centering
  \includegraphics[width=2.4in]{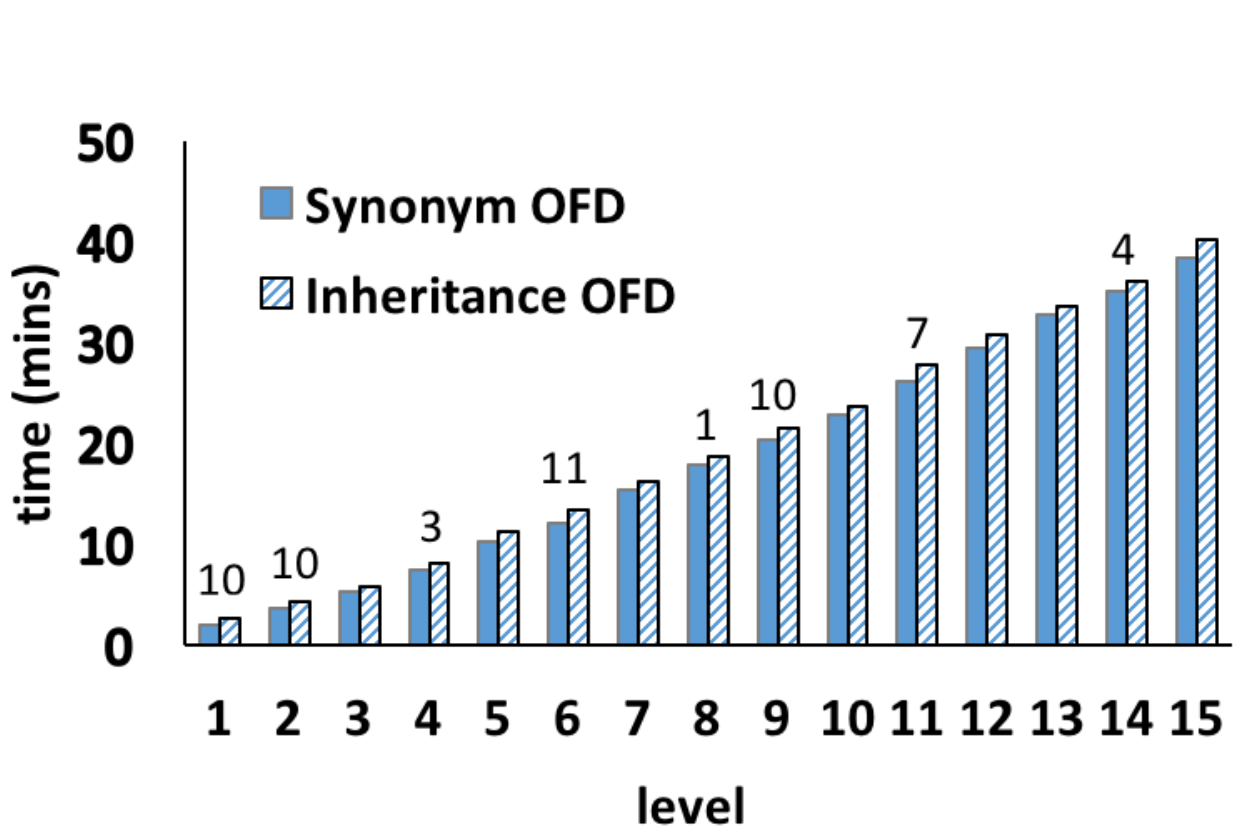}
\caption{Runtime per level. \label{fig:level}}
	\vspace{-0.1cm}
   \endminipage\hfill
      \minipage[b]{0.5\linewidth}
   \centering
  \includegraphics[width=2.4in]{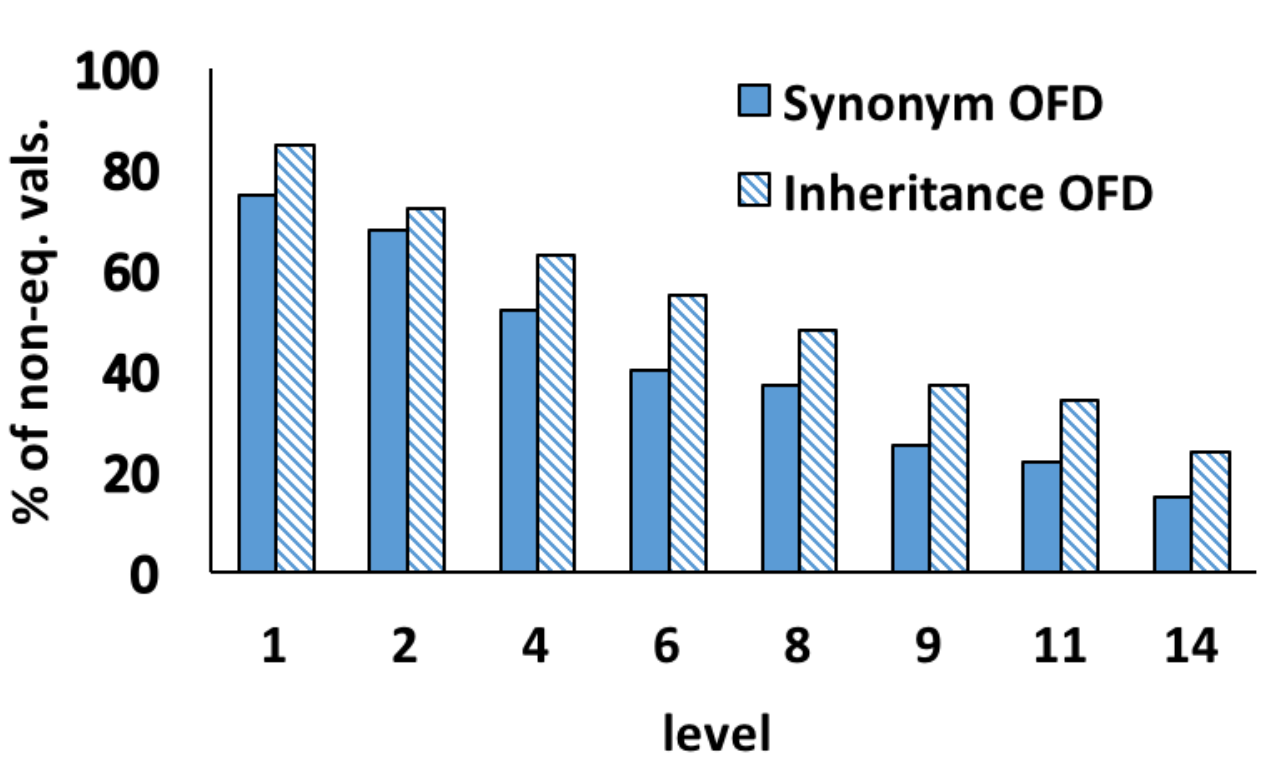}
 \vspace{-0.3cm}
  \caption{Synonym and inheritance values found per level.}\label{fig:ofdclean1}
  \vspace{-0.3cm}
   \endminipage
   \minipage[b]{0.5\linewidth}
   \centering
  \includegraphics[width=2.4in]{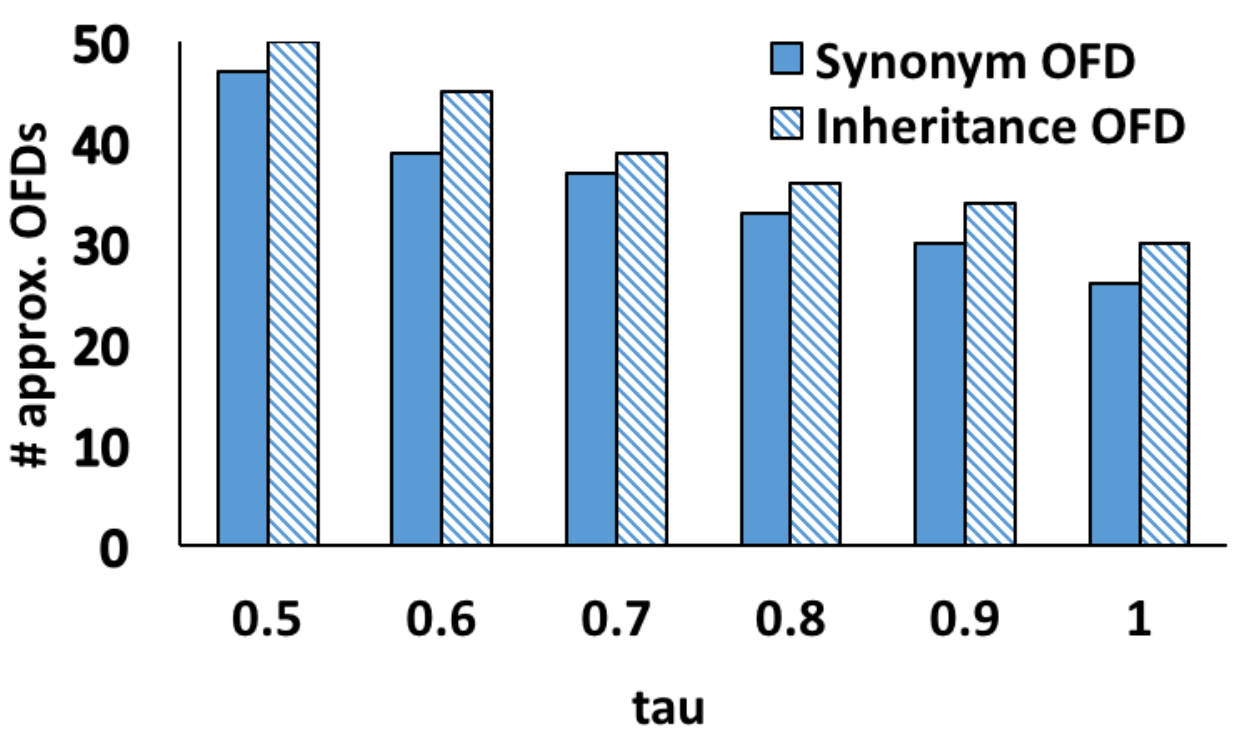}
 \vspace{-0.3cm}
  \caption{Number of approximate OFDs vs. support.}\label{fig:ofdclean2}
  \vspace{-0.3cm}
   \endminipage\hfill

\end{figure*}


\subsection{Optimization Benefits}
\label{sec:opt}
We now evaluate the efficiency of our optimizations using the clinical trials dataset.  We focus our evaluation on \emph{Optimizations 2, 3} and \emph{4 (Opt-2, Opt-3, Opt-4)}. Recall that for $O: X \rightarrow A$, \emph{Optimization 1} prunes trivial dependencies where $A \in X$.   Our current lattice based search does not compose candidates where $A \in X$, so we do not generate trivial dependencies.   \emph{Opt-2} prunes candidates containing a superset of $X$ if $O$ already holds over $I$.  \emph{Opt-3} and \emph{Opt-4} eliminate the verification step to check whether $O$ is an OFD, if $X$ is a key or $O$ is found to be a traditional FD, respectively.  We use 500K tuples from the clinical trials dataset.  Our optimizations achieve significant performance improvements.  For instance, at 100K tuples for Opt-4, we are able to reduce running times by 59\%.

\textbf{Experiment-3: Pruning Non-Minimal OFDs (Opt-2).}  Figure \ref{fig:opt2} shows the running times of our discovery algorithm with and without Opt-2 implemented, labeled as `FASTOFD\_optimized' and `FASTOFD', respectively.  Opt-2 achieves an average 31.4\% improvement in running time, due to the aggressive pruning of redundant candidates at the lower lattice levels.

\textbf{Experiment-4: Exploiting Keys (Opt-3).} Figure \ref{fig:opt3} shows the running times to evaluate Opt-3.  As expected, we see a smaller performance improvement (compared to Opt-2) since we reduce the verification time due to existence of keys, rather than pruning candidates.  We found two key attributes in the clinical data: (1) \texttt{NCTID} representing the clinical trials.gov unique identifier for a clinical study; and (2) \texttt{OrgStudyID} an identifier for a group of related studies.  Opt-3 achieves an average 14.3\% improvement in running time.  Given the existence of more keys, we would expect to see a greater improvement in running time.  

\textbf{Experiment-5: Exploiting FDs (Opt-4).} Similarly, Figure \ref{fig:opt4} shows the evaluation for Opt-4.  Since the original dataset did not have any FDs that were satisfied over the entire relation,  we modified the data to include five FDs: 
\begin{itemize}
\item	F1: overall\_status $\rightarrow$ number\_of\_groups, 
\item	F2: study\_design $\rightarrow$ study\_type,
\item    F3: [condition, time\_frame] $\rightarrow$ measure,
\item   F4: [safety\_issue, study\_type] $\rightarrow$ eligibility\_minimum\_age, 		
\item	F5: [condition, country] $\rightarrow$ drug\_name.
\end{itemize}

Given this set of FDs, the running times decrease by an average of 27\%.  We expect that the performance gains are proportional to the number of satisfying FDs in the data; an increased number of satisfying FDs leads to less time spent verifying candidate OFDs.  Overall, our optimizations are able to achieve an average of 24\% improvement in running time, and our canonical representations are effective in avoiding redundancy.

\subsection{Efficiency}
\textbf{Experiment-6: Effectiveness over lattice levels.} 
We argue that compact OFDs (those involving a small number of attributes) are the most interesting.  OFDs with a larger number of attributes contain more unique equivalence classes.  Thus, a less compact dependency may hold over the relation, but not be very meaningful due to overfitting.  We evaluate the efficiency of our techniques to discover these compact dependencies.  To do this, we measure the number of OFDs, and the time spent, at each level of the lattice using the clinical trials data.  OFDs discovered at the upper levels of the lattice (involving fewer attributes) are more desirable than those discovered at the lower levels.  Figure \ref{fig:level} shows the running time at each level of the lattice, and the total number of OFDs found at each level.  Table \ref{tab:countlvl} shows the breakdown for each type of OFD per level (missing levels indicate no OFDs found).

For synonym OFDs, approximately 61\% are found in the first 6 levels taking about 25\% of the total time.  For inheritance OFDs, the results are a bit better, where 63\% of the dependencies are found in the first 6 levels taking 16\% of the total time.  The remaining 35-40\% of dependencies (found in the lower levels) are not as compact, and the time to discover these OFDs would take well over 70\% of the total time.   Since most of the interesting OFDs are found at the top levels, we can prune the lower levels (beyond a threshold) to improve overall running times.

\begin{table}
  \centering
    \begin{tabular}{|c||c|c|}
       \hline
        level & synonym OFDs & inheritance OFDs \\
       \hline \hline
       1 & 5   & 5   \\
       2 & 4  & 6  \\
       4 & 3  & - \\
       6 & 4 & 7 \\
       8 & - & 1 \\
       9 & 5 & 5 \\
       11 & 3 & 4 \\
       14 & 2 & 2 \\
       \hline
   \end{tabular}
    \captionof{table}{Number of OFDs found per level.}
    \vskip -0.9cm
 \label{tab:countlvl}
 \end{table}

\subsection{Qualitative Evaluation}
\textbf{Experiment-7: Evaluation of OFD Accuracy.} To measure the accuracy of the discovered OFDs, we used clean versions of the data containing a known number of satisfying synonym and inheritance OFDs as the ground truth.  We then randomly injected 2\% error, and measured the precision and recall.  We found the discovered dependencies to be correct, and the recall values ranged between 70\%-80\%.  Figure \ref{fig:recall} shows the recall results for the clinical trials and census datasets.  We found the discovered OFDs to be intuitive and relevant.  For example, in the census data, synonym OFD $O_{1}$: OCCUP $\rightarrow_{syn}$ SAL states that equivalent jobs earn a similar salary (with SAL being a categorical attribute representing salary levels).   In the clinical data, two examples are the inheritance OFD $O_{2}$: [SYMP, DIAG] $\rightarrow_{\theta}$ DRUG, and the synonym OFD $O_{3}$: CONDITION $\rightarrow_{syn}$ DRUG.  In $O_{2}$, a set of symptoms and diagnosis leads to a prescribed treatment from the same drug family; e.g., a headache diagnosed as a migraine is prescribed ibuprofen (which is-a non-steroid anti-inflammatory drug) for pain relief.  In $O_{3}$, we found that specific medical conditions are treated with similar drugs, but the drug names vary across countries.  For example, \emph{fever} is treated with \emph{acetaminophen} in Canada, and with \emph{Paracetamol} in India.  The discovered dependencies re-affirm that many applications contain domain specific synonym and inheritance attribute relationships that go beyond equality, which are not currently captured by existing techniques. 



\subsection{Data Cleaning with OFDs}
\label{sec:ofdclean}
We now evaluate the utility of OFDs for data cleaning using the clinical data.  OFDs provide rich contextual information that is used to distinguish synonymous and inheritance values that are not syntactically equivalent but refer to the same entity.  In addition, approximate OFDs can identify potentially dirty data values, and provide recommendations of possible correct values for repair. 

\textbf{Experiment-8: Evaluation of Synonyms and Inheritance Values.}
To quantify the benefits of OFDs versus traditional FDs, which are commonly used in data cleaning techniques, we aim to capture the value that OFDs provide, namely, the number of syntactically `non-equal' values referring to the same entity.   For example, under traditional FDs, a dependency \texttt{CTRY} $\rightarrow$ \texttt{CC}, with CTRY value `Canada' mapping to CC values `CA', `CAN', and `CAD', would all be considered as errors.  However, under a synonym OFD, these tuples are considered clean, since `CA', `CAN', and `CAD' are acceptable representations of `Canada'.   

Table \ref{tab:countlvl} shows the number of synonym and inheritance FDs found at each level in the clinical data.   For each discovered OFD, we sample and compute the percentage of tuples containing syntactically non-equal values in the consequent.  These values represent either synonyms or values participating in an inheritance relationship.  Under FD based data cleaning, these tuples would be considered errors.  However, by using OFDs, these false positive errors are saved since they are not true errors.  Figure \ref{fig:ofdclean1} shows the percentage of synonym and inheritance values at each lattice level where an OFD was found.  Overall, we observe that a significantly large percentage of tuples are falsely considered as errors under a traditional FD based data cleaning model.  For example, at level 1, 75\% and 84\% of the values in synonym and inheritance OFDs, respectively, contain non-equal values.  Only after level 6, we see that the number of tuples containing equal values in the consequent comprise over half of the satisfying tuples in the OFD.  This is somewhat expected, since the dependency becomes more constrained with an increasing number of antecedent attributes.  By correctly recognizing these `erroneous' tuples as clean, we reduce the computational effort of existing data cleaning techniques, and the manual burden for users to decipher through these falsely categorized errors.  

\textbf{Experiment-9: Evaluation of Approximate OFDs.}
 Approximate OFDs can be used in data cleaning to identify potentially dirty data values, and recommend values (from the satisfying set of tuples) for data repair.  Figure \ref{fig:ofdclean2} shows the number of discovered synonym and inheritance OFDs for values of $\tau \geq 0.5$ over the clinical data.  As expected, we observe a moderate increase in the number of discovered dependencies as the support level decreases.  
 
 We examined a sample of the approximate OFDs at each $\tau$ level, and make two observations.  First, data repairs for the `dirty' tuples  at $\tau = 0.8$ and $\tau = 0.9$ were more easily determined than at lower $\tau$ levels due to the increased support of satisfying tuples, since patterns of values occurred with higher frequency (using a frequency based repair model).  For example, in approximate synonym OFD [CC, DIAG] $\rightarrow_{syn}$ DRUG, we found that references to drug names fluctuated greatly depending on the country, the diagnosis, and whether the drug is referred to as a brand name or in its natural (biological) form.  For example, a medication commonly used to treat anemia is also used to treat cancer and kidney failure, and is referred to by at least four different names: in its natural form as `Erythropoietin', and synthetically as `Epoetin', `Epogen', `Procrit', and `Darbepoetin Alfa'.   Secondly, approximate OFDs at $\tau = 0.5$ and $\tau = 0.6$ appeared to suffer from the expected overfitting problem.  At these levels, due to many lower frequency tuple pattern values, it is not as clear as to how the dirty tuples should be repaired.  
 
 In summary, approximate OFDs serve as a new data quality rule to capture domain relationships that were not considered by existing data quality dependencies.  Our observations indicate that approximate OFDs with higher support contain more reliable values for repairing inconsistent, and potentially dirty values.  As next steps, we intend to explore how approximate OFDs can be incorporated into automated data cleaning techniques to provide richer semantics during the data cleaning process.

\section{Related Work}
\label{sec:rw}

\textbf{Dependency discovery.}  Previous work on extracting dependencies from data includes
discovery of functional dependencies (FDs) \cite{HKPT98,LPL00,WGR01,NC01,PPE+15}, conditional functional dependencies (CFDs) \cite{CM08,GKKSY08,FGLX11}, inclusion dependencies \cite{PKQJN15}, order dependencies \cite{ODTR16}, matching dependencies \cite{SC09}, and denial constraints \cite{CIP13}.  In previous FD discovery algorithms, TANE \cite{HKPT98} and DepMiner \cite{LPL00} search the attribute lattice in a level-wise manner for a minimal FD cover.   In CFD discovery algorithms, a similar lattice traversal is used to identify a subset of tuples that functionally hold over a relational instance \cite{CM08,FGLX11}.  In our work, we generalize the lattice based level-wise strategy for discovering synonym and inheritance OFDs.

Previous work has extended classical FDs to consider attribute domains that contain a partial order, and to support time-related dependencies in temporal databases \cite{WNC99,JSS96,WBBJ97,Wijsen99}.   Wijsen et al.\ propose \emph{Roll-Up Dependencies (RUDs)} that generalize FDs for attribute domains containing concept hierarchies that are commonly found in data mining and OLAP applications \cite{WNC99}.  For example, a \texttt{time} attribute contains values that can be organized into a partial order, measured in days, weeks, months, etc.  RUDs capture roll-up semantics from one or more attributes that have been aggregated at finer levels.  The set of possible generalizations for an attribute set in a candidate RUD is modelled as a lattice. Similar to our approach, the RUD discovery algorithm traverses the lattice in a levelwise top-down manner.   However, our inheritance OFDs, in particular, capture containment semantics (similar to is-a semantics) that are not modelled by RUDs.

Similar to traditional FDs, Jensen et al.\ propose \emph{Temporal FDs (TFDs)} that hold over a snapshot of a temporal database, called a timeslice \cite{JSS96}.  Extensions of TFDs include constraining tuples across multiple timeslices \cite{WBBJ97}, and generalizing the time model to include objects and classes, whereby an attribute value is no longer necessarily atomic, but may refer to an object of another class \cite{Wijsen99}.  This object referencing implicitly provides referential integrity, but does not consider the synonym and inheritance relationships we investigate in our work.   We focus on identifying synonym and inheritance dependencies containing atomic data types at a given timeslice. In the future, it will be interesting to consider how our dependencies can be extended to include object classes, and their variation over time.

\textbf{Data Cleaning.}  Constraint based data cleaning solutions have used FDs as a benchmark to  propose data repairs such that the data and the FDs are consistent \cite{BFFR05}.  Recent extensions have relaxed the notion of equality in FDs, and used similarity and matching functions to identify errors, and propose repairs \cite{BKL11,HTL+17}.  While our work is in similar spirit,  we differ in the following ways: (1) the similarity functions match values based on syntactic string similarity, where functions such as edit-distance, Jaccard, and Euclidean are used.  Hence, values such as  'India' and 'Bharat' would be dissimilar; (2) the notion of senses is not considered, and  similarity under multiple interpretations is not possible;  and (3)  inheritance (is-a) relationships  are not supported in existing techniques.

\textbf{Integrity constraints on ontologies and graphs.} Ontologies are used to model concepts, entities, and relationships for a given domain.
Existing techniques have proposed FDs over RDF triples based on the co-occurrence of values.  However, these FDs do not consider structural requirements to specify which entities should carry the values \cite{ACP10,HGPW16}.   Motik et al.\ define integrity constraints using the Web Ontology Language (OWL).  OWL ontologies are often incomplete, whereas many databases in practice are complete \cite{MHS07}.  They propose an extension of OWL with integrity constraints to validate completeness in the ontology by defining inclusion dependencies and domain constraints to check for missing values and valid domain values within an ontology.  The proposed constraints do not model functional dependencies (as proposed in our work) since the focus is on data completeness.  Furthermore, these existing techniques do not consider the notion of senses to distinguish similar terms under an interpretation.  

Fan et. al., define keys for graphs based on patterns that specify topological constraints and value bindings to perform entity matching \cite{FFTD15}.  Keys contain variables that are bound to constant values satisfying node and value equality.  The authors focus on the definition of keys (not their discovery), and present three sub-graph entity matching algorithms that utilize keys.  In subsequent work, Fan et. al.\ propose functional dependencies for graphs (GFDs) since FDs cannot be expressed via keys \cite{FWX16}.  GFDs contain topological constraints to identify the entities participating in the dependency, and value bindings (similar to conditional FDs) that specify dependencies among the attribute values.  GFDs model is-a relationships (e.g., $y$ is-a $x$) by assuming this inheritance relationship is known in advance, and then enforce the requirement that for any property $A$ of $x$ must also be true for $y$, i.e., $x.A = y.A$.  In our work, we focus on the \emph{discovery} of is-a relationships where the left-hand-side attribute values determine inheritance relationships between right-hand-side attribute values.  
While our work is similar in spirit, we identify attribute relationships that go beyond equality (i.e., synonyms and inheritance).  In contrast to keys, our discovered dependencies are value based (no variables are present).  In our work, we consider the notion of senses that states how a dependency should be interpreted, since multiple interpretations are possible for a given ontology; these interpretations are not considered in existing techniques.  Lastly,  we study the axiomatization and inference of synonym and inheritance relationships in OFDs, which were not studied in previous work.
\vspace{-0.1cm}


\section{Conclusions}
\label{sec:conclusion}

We present a new class of data quality rules, Ontology Functional Dependencies (OFDs) that capture domain relationships found in ontologies, namely synonyms and inheritance.  We develop an axiomatization and inference procedure for OFDs, and design an OFD discovery algorithm that identifies a minimal set of OFDs.  
Our evaluation showed that our algorithm scales well, discovers compact dependencies, and our axiom-driven optimizations achieve significant performance improvements.   Finally, we showed that OFDs are a useful data quality rule to capture domain relationships, and significantly reduces the number of false positive errors in data cleaning solutions that rely on traditional FDs.   In  future work, we intend to consider extensions to other relationships such as \emph{component-of},  and the use of ontologies to discover other types of data quality rules such as conditional FDs and denial constraints.   
\vspace{-0.1cm}




\small
\bibliographystyle{abbrv}
\bibliography{rw}
\balance

\eat{
\newpage
}


\end{document}